\pdfoutput=1
\documentclass[letterpaper,10pt,twocolumn]{article}
\usepackage[T1]{fontenc} 
\usepackage{graphicx}
\usepackage[font=small]{caption}
\usepackage{float}
\usepackage{hyperref}
\usepackage{booktabs,makecell} 
\usepackage[numbers]{natbib} 
\usepackage{amsmath}
\usepackage{enumitem}
\usepackage{algorithm}
\usepackage[noend]{algpseudocode}
\usepackage{mathptmx}
\usepackage{authblk}
\setlist[itemize]{leftmargin=10pt}

\algrenewcommand\textproc{} 
\begin{document}

\title{Wormhole: A Fast Ordered Index for In-memory Data Management}

\author[$\dag$]{Xingbo Wu}
\author[$\ddag$]{Fan Ni}
\author[$\ddag$]{Song Jiang}
\affil[$\dag$]{University of Illinois at Chicago}
\affil[$\ddag$]{University of Texas at Arlington}

\date{}
\maketitle

\begin{abstract}

In-memory data management systems, such as key-value stores,
have become an essential infrastructure in today's big-data processing and cloud computing.
They rely on efficient index structures to access data.
While unordered indexes, such as hash tables, can perform point search with $O(1)$ time,
they cannot be used in many scenarios where range queries must be supported.
Many ordered indexes, such as B+ tree and skip list,
have a $O(\log N)$ lookup cost, where $N$ is number of keys in an index.
For an ordered index hosting billions of keys,
it may take more than 30 key-comparisons in a lookup,
which is an order of magnitude more expensive than that on a hash table.
With availability of large memory and fast network in today's data centers,
this $O(\log N)$ time is taking a heavy toll on applications that rely on ordered indexes.

In this paper we introduce a new ordered index structure, named \textit{Wormhole},
that takes $O(\log L)$ worst-case time for looking up a key with a length of $L$.
The low cost is achieved by simultaneously leveraging strengths of three indexing structures,
namely hash table, prefix tree, and B+ tree, to orchestrate a single fast ordered index.
Wormhole's range operations can be performed by a linear scan of a list after an initial lookup.
This improvement of access efficiency does not come at a price of compromised space efficiency.
Instead, Wormhole's index space is comparable to those of B+ tree and skip list.
Experiment results show that Wormhole outperforms skip list, B+ tree, ART, and Masstree
by up to 8.4$\times$, 4.9$\times$, 4.3$\times$, and 6.6$\times$ in terms of key lookup throughput, respectively.

\end{abstract}

\section{Introduction}
\label{sec:wh-intro}

A common approach to building a high-performance data management system is to host all of its data
and metadata in the main memory~\cite{MemSQL,SQLite,Redis,lmdb}.
However, when expensive I/O operations are removed (at least from the critical path),
index operations become a major source of the system's cost, reportedly consuming
14--94\% of query execution time in today's in-memory databases~\cite{KGP13}.
Recent studies have proposed many optimizations to improve them with a major focus on
hash-table-based key-value (KV) systems, including efforts on avoiding chaining in hash tables,
improving memory access through cache prefetching,
and exploiting parallelism with fine-grained locking~\cite{FAK13,LLL16,NFG13}.
With these efforts the performance of index lookup can be pushed close to the hardware's limit,
where each lookup needs only one or two memory accesses to reach the requested data~\cite{LLL16}.

Unfortunately, the $O(1)$ lookup performance and benefits of the optimizations are not available
to ordered indexes used in important applications, such as B+ tree in LMDB~\cite{lmdb},
and skip list in LevelDB~\cite{leveldb}. Ordered indexes are required to support
range operations,
though the indexes can be (much) more expensive than hash tables supporting only point operations.
Example range operations include searching for all keys in a given key range or for keys of a common
prefix. It has been proved that lookup cost in a comparison-based ordered index
is $O(\log N)$ key comparisons,
where $N$ is the number of keys in the index~\cite{CLR09}.
As an example, in a B+ tree of one million keys
a lookup requires about 20 key comparisons on average.
When the B+ tree grows to billions of keys,
which is not rare with small KV items managed in today's large memory of hundreds of GBs,
on average 30 or more key-comparisons are required for a lookup.
Lookup in both examples can be an order of magnitude slower than that in hash tables.
Furthermore, searching in a big index with large footprints increases working
set size and makes CPU cache less effective.
While nodes in the index are usually linked by pointers,
pointer chasing is a common access pattern in the search operations.
Therefore, excessive cache and TLB misses may incur tens of DRAM accesses
in a lookup operation~\cite{WNJ17}.
The performance gap between ordered and unordered indexes has been significantly widened.
As a result, improving ordered indexes to support efficient search operations
has become increasingly important.

As a potential solution to reduce the search overhead,
prefix tree, also known as trie,
may be adopted as an ordered index, where a key's location is solely determined by the key's content
(a string of \textit{tokens}, e.g., a byte string), rather than by the key's relative order
in the entire keyset. Accordingly, trie's search cost is determined by the number of tokens
in the search key ($L$), instead of the number of keys ($N$) in the index.
This unique feature makes it possible for tries to perform search faster than the
comparison-based ordered indexes, such as B+ tree and skip list. As an example,
for a trie where keys are 4-byte integers and each byte is a token,
the search cost is upper-bounded by a constant ($4$) regardless of the number of keys in the index.
This makes trie favorable in workloads dominated by short keys,
such as searching in IPv4 routing tables where all of the keys are 32-bit integers.
However, if keys are long (e.g., URLs of tens of bytes long),
even with a small set of keys in the trie, the search cost can be consistently high
(possibly substantially higher than the $O(\log N)$ cost in other indexes).
As reported in a study of Facebook's KV cache workloads on its production Memcached system,
most keys have a size between 20 to 40 bytes~\cite{AXF12}, which makes trie an undesirable choice.
It is noted that the \textit{path compression} technique may help to reduce
a trie's search cost~\cite{LKN13}.
However, its efficacy highly depends on the key contents,
and there is no assurance that its $O(L)$ cost can always be reduced.
Together with its issues of inflated index size and fragmented memory usage~\cite{LKN13},
trie has not been an index structure of choice in general-purpose in-memory data management systems.

In this paper we propose a new ordered index structure, named \textit{Wormhole},
to bridge the performance gap between hash tables and ordered indexes
for high-performance in-memory data management.
Wormhole efficiently supports all common index operations,
including lookup, insertion, deletion, and range query.
Wormhole has a lookup cost of $O(\log L)$ memory accesses,
where $L$ is the length of search key (actual number of accesses can be (much) smaller than $\log_2 L$).
With a reasonably bounded key length (e.g., 1000 bytes), the cost can be considered as $O(1)$,
much lower than that of other ordered indexes, especially for a very-large-scale KV store.
In addition to lookup, other operations, such as insertion, deletion, and range query, are also efficiently supported.
In the meantime,
Wormhole has a space cost comparable to B+ tree, and often much lower than trie.

This improvement is achieved by leveraging strengths of three data structures,
namely, space efficiency of B+ tree (by storing multiple items in a tree node),
trie's search time independent of store size, and hash-table's $O(1)$ search time,
to orchestrate a single efficient index. Specifically, we first use a trie structure
to replace the non-leaf section of a B+ tree structure in order to remove the $N$ factor
in the B+ tree's $O(\log N)$ search time.
We then use a hash table to reduce the lookup cost on the trie structure
to $O(\log L)$, where $L$ is the search key length.
We further apply various optimizations in the new structure to realize
its full performance potential and maximize its measurable performance.
The proposed ordered index is named \textit{Wormhole}
for its capability of jumping on the search path from the tree root to a leaf node.

We design and implement an in-memory Wormhole index and extensively
evaluate it in comparison with several representative indexes,
including B+ tree, skip list, Adaptive Radix Tree (ART)~\cite{LKN13},
and Masstree (a highly optimized trie-like index)~\cite{MKM12}.
Experiment results show that Wormhole outperforms these indexes
by up to 8.4$\times$, 4.9$\times$, 4.3$\times$, and 6.6$\times$,
in terms of key lookup throughput, respectively.
We also compare Wormhole with a highly optimized Cuckoo hash table when range queries
are not required. The results show that Wormhole achieves point-lookup throughput
30--92\% of the hash-table's throughput.

The rest of this paper is organized as below.
Section~\ref{sec:wh-index} introduces design of Wormhole's core data structure.
Section~\ref{sec:wh-opt} describes techniques for efficient implementation of the Wormhole index.
Section~\ref{sec:wh-eval} presents experiment setup, workloads, and evaluation results.
Section~\ref{sec:wh-related} discusses the related work, and
Section~\ref{sec:wh-conc} concludes.

\section{The Wormhole Data Structure}
\label{sec:wh-index}

In this section we introduce the Wormhole index structure,
which has significantly lower asymptotic lookup time than existing ordered indexes,
without increasing demand on space and cost of other modification operations, such as insertion.
To help understand how Wormhole achieves this improvement,
we start from B+ tree and progressively evolve it to the structure of Wormhole.

\subsection{Background: Lookup in the B+ Tree}
\label{sec:wh-bt}

\begin{figure}[!t]
  \centering
  \includegraphics[width=0.7\columnwidth]{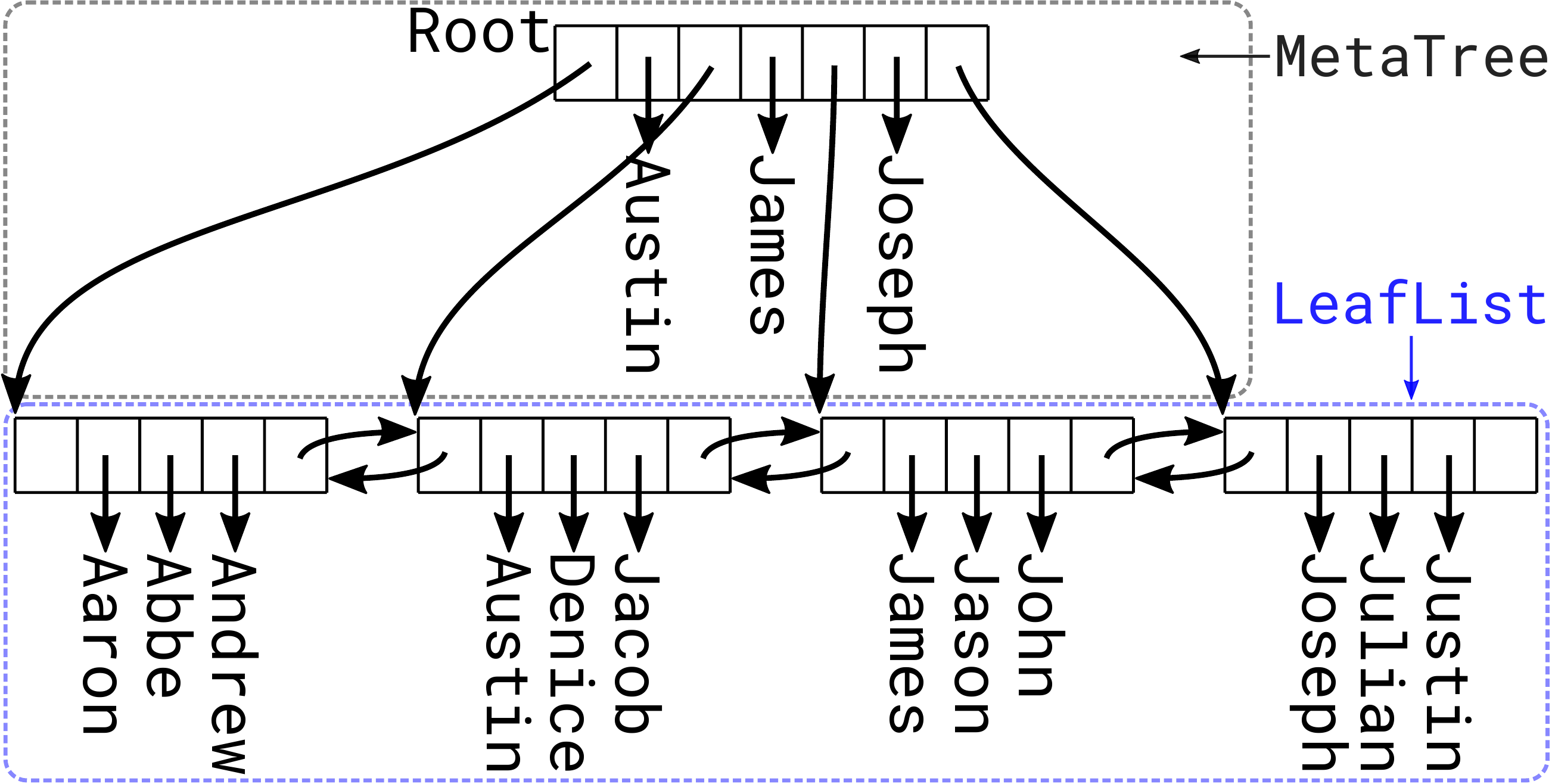}
  \caption{An example B+ tree containing 12 keys}
  \label{fig:arch1}
\end{figure}

Figure~\ref{fig:arch1} shows a small set of 12 keys indexed in an example B+
tree, where each character is a token. While a key in the index is usually associated with a value,
we omit the values in the discussion and only use keys to represent
KV items to focus on time and space costs of index
operations. The example B+ tree has one internal node (the root
node) and four leaf nodes. In the B+ tree all keys are placed in leaf
nodes while internal nodes store a subset of the keys to facilitate
locating search keys at leaf nodes.
Keys in a leaf node are usually sorted and all leaf
nodes are often linked into a fully sorted list to support range
operations with a linear scan on it. We name the sorted list
\textit{LeafList}, and the remaining structure of the index as
\textit{MetaTree}, as shown in Figure~\ref{fig:arch1}.

MetaTree is used to accelerate the process of
locating a leaf node that potentially stores a given
search key. A search within the leaf node is conducted
thereafter. Because a leaf node's size, or number of keys held in the
node, is bounded in a predefined range $[\lceil \frac{k}{2}\rceil, k]$ ($k$ is a predefined constant integer),
the search with a leaf node takes $O(1)$ time.
Accordingly, the major search cost is incurred in the MetaTree, which is $\log_2 \frac{N}{k}$
or $O(\log N)$ (N is the number of indexed keys).
As the B+ tree grows, the MetaTree will contain more levels of internal nodes,
and the search cost will increase at a rate of $O(\log N)$.
Our first design effort is to replace the MetaTree with a structure whose search cost is not tied to $N$.

\subsection{Replacing the MetaTree with a Trie}
\label{sec:wh-trie}

\begin{figure}[!t]
  \centering
  \includegraphics[width=0.69\columnwidth]{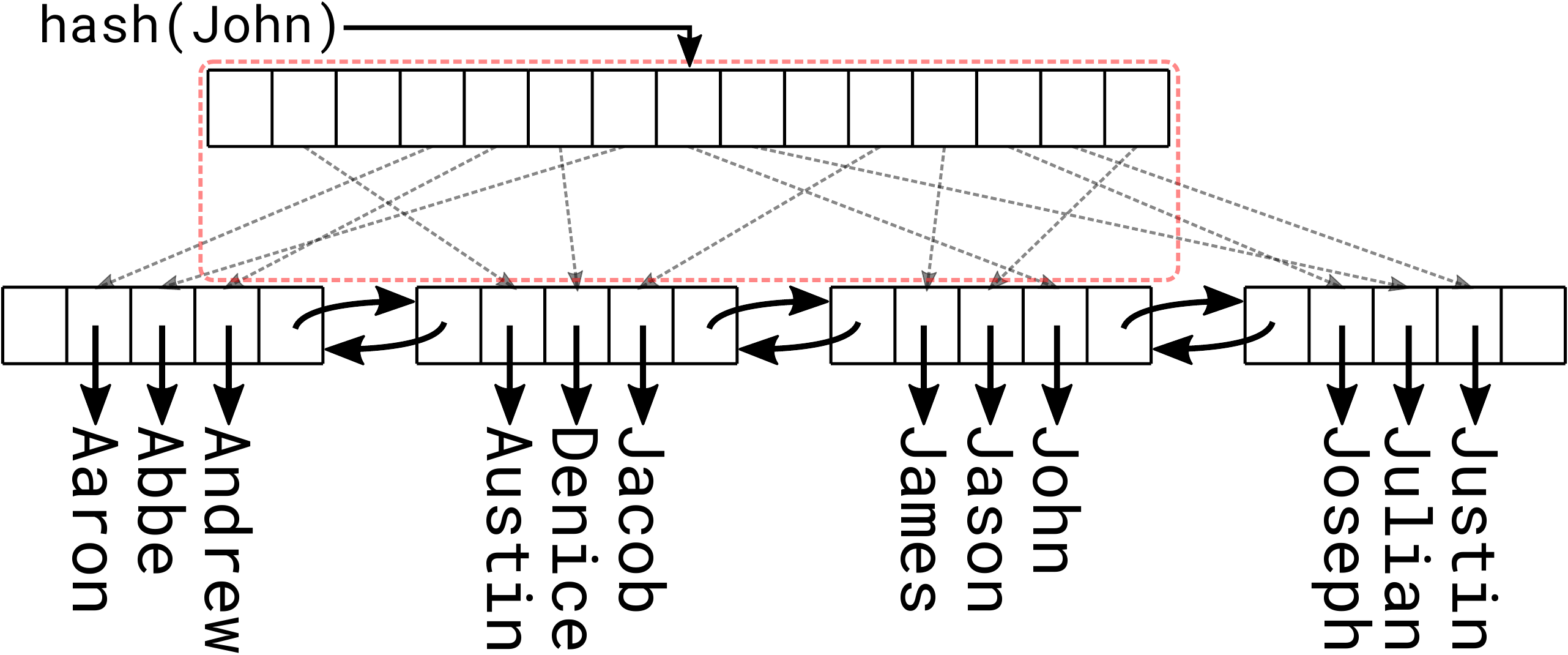}
  \caption{Replacing B+ tree's \textit{MetaTree} with hash table}
  \label{fig:archx}
\end{figure}

An intuitive idea on B+ tree's improvement is to replace its MetaTree structure with a hash table,
as illustrated in Figure~\ref{fig:archx}. This can reduce the search cost to $O(1)$. However, this
use of hash table does not support inserting a new key at the correct position in the sorted LeafList.
It also does not support range queries whose search identifiers are not existent in the index,
such as search for keys between ``Brown'' and ``John'' or search for keys with a prefix of ``J''
in the example index shown in Figure~\ref{fig:archx}, where ``Brown'' and ``J'' are not in the index.
Therefore, the MetaTree itself must organize keys in an ordered fashion.
Another issue is that the hash table requires an entry (or pointer) for every key in the index,
demanding a space cost higher than MetaTree.

\begin{figure}[!t]
  \centering
  \includegraphics[width=0.7\columnwidth]{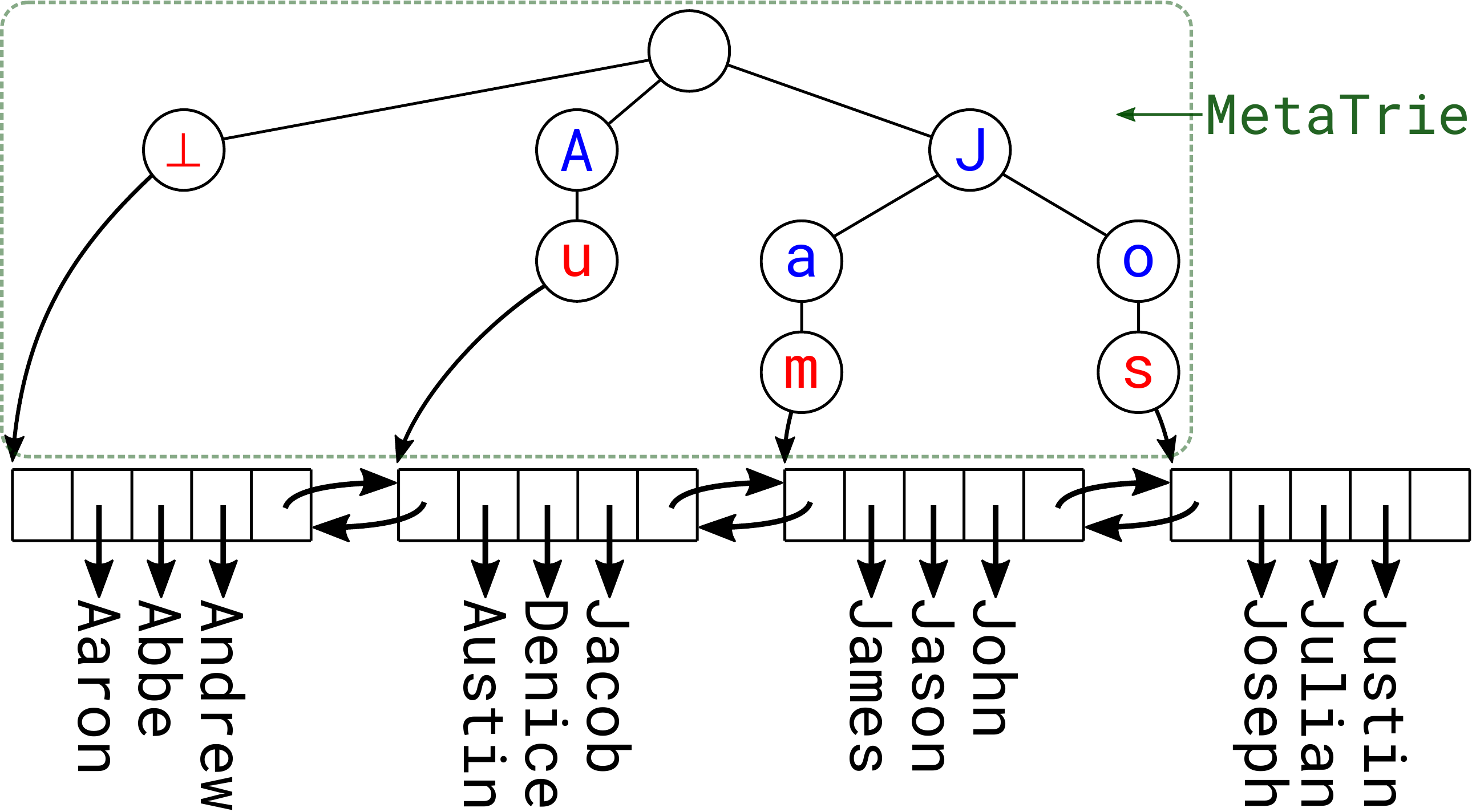}
  \caption{Replacing B+ tree's \textit{MetaTree} with \textit{MetaTrie}}
  \label{fig:arch2}
\end{figure}

To address the issues, trie can be a better replacement as it is an ordered index and its lookup cost
($O(L)$, where $L$ is the search key length) is not tied to $N$, the number of keys in the index.
Figure~\ref{fig:arch2} illustrates the index evolved from B+ tree
with its MetaTree structure replaced by a trie structure named \textit{MetaTrie}.
For each node in the LeafList we create a key as its \textit{anchor} and insert it into MetaTrie.
A node's anchor key is to serve as a borderline between this node and the node immediately
on its left, assuming the sorted LeafList is laid out horizontally in an ascending order
as shown in Figure~\ref{fig:arch2}. Specifically, the anchor key (\textit{anchor-key}) of a node (Node$_b$),
must meet the following two conditions:

\begin{itemize}[itemsep=0ex,topsep=0ex,partopsep=0ex,parsep=0ex]
\item \textbf{The Ordering Condition}: $\textit{left-key}<\textit{anchor-key}\leq\textit{node-key}$,
  where \textit{left-key} represents any key in the node (Node$_a$) immediately left to Node$_b$,
  and \textit{node-key} represents any key in Node$_b$.
  If Node$_b$ is the left-most node in the LeafList,
  the condition is $\textit{anchor-key}\leq\textit{node-key}$.
\item \textbf{The Prefix Condition}: An anchor key cannot be a prefix of another anchor key.
\end{itemize}

When an anchor key is inserted into the MetaTrie,
one new leaf node  corresponding to the key is created in the trie. In addition, any prefix of the key is inserted to the trie as its internal node, assuming it is not yet in the trie.
We use the prefix condition to make sure every anchor key has a corresponding leaf node in the MetaTrie. 

In the formation of an anchor key, we aim to minimize the key length to reduce the MetaTrie size.
To this end we design a method to form an anchor key for the aforementioned Node$_b$ in compliance
with the two conditions, assuming the smallest token, denoted by $\bot$,
does not appear in regular keys (other than the anchor keys) on the LeafList.
We will remove the restriction on use of the smallest token later.
We denote the smallest key in Node$_b$ as $\langle P_1P_2...P_kB_1B_2...B_m\rangle$
and the largest key in Node$_a$ as $\langle P_1P_2...P_kA_1A_2...A_n\rangle$,
and $A_1 < B_1$, where $P_i$ ($1\leq i\leq k$), $A_i$ ($1\leq i\leq m$),
and $B_i$ ($1\leq i\leq n$) represent the keys' tokens. If $k$ or $n$ is 0, it represents the corresponding key segment does not exist. 
Accordingly, $\langle P_1P_2...P_k\rangle$ is the longest common prefix of the two keys.
Assuming Node$_b$ is a new leaf node whose anchor key has not been determined, we form its anchor key as follows:

\begin{itemize}[itemsep=0ex,topsep=0ex,partopsep=0ex,parsep=0ex]
\item If Node$_b$ is not the left-most node on the LeafList ($m>0$),
we will check whether $\langle P_1P_2...P_kB_1\rangle$
is a prefix of the anchor key of the node immediately after
Node$_b$ on the LeafList, denoted Node$_c$.\footnote{Note
that if $\langle P_1P_2...P_kB_1\rangle$ is a prefix of any other anchor,
it must be a prefix of Node$_c$'s anchor.}
If not (including the case where Node$_c$ does not exist),
Node$_b$'s anchor is $\langle P_1P_2...P_kB_1\rangle$.
Otherwise,
Node$_b$'s anchor is $\langle P_1P_2...P_kB_1\bot\rangle$,
which cannot be a prefix of Node$_c$'s anchor.
This is because a $(k+2)$\textsuperscript{th} token of Node$_c$'s anchor key must be larger than $\bot$.
We then check whether Node$_a$'s anchor is a prefix of Node$_b$'s anchor
(Node$_a$ is $\langle P_1P_2...P_j\rangle$, where $j \leq k$).
If so, Node$_a$'s anchor will be changed to $\langle P_1P_2...P_j\bot\rangle$. Note that by appending the $\bot$ token to meet the anchor key's prefix condition, its ordering condition can be violated. To accommodate the situation, the $\bot$ is ignored in the ordering condition test without compromising the correctness.   

\item Otherwise (Node$_b$ is the left-most node), its anchor is $\bot$,
which is not any other anchor's prefix.

\end{itemize}

Using the method the four leaf nodes in Figure~\ref{fig:arch2}, starting from the left-most one,
have their respective anchors as ``$\bot$'', ``$Au$'', ``$Jam$'', and ``$Jos$''.
All the anchors and their prefixes are inserted into the MetaTrie.

\subsection{Performing Search on MetaTrie}
\label{sec:trie1-lookup}

The basic lookup operation on the MetaTrie with a search key is similar to that in a conventional
trie structure, which is to match tokens in the key to those in the trie
one at a time and walk down the trie level by level accordingly. If the search key is ``Joseph'' 
in the example index shown in Figure~\ref{fig:arch2}, it will match the anchor key ``Jos'',
which leads the lookup to the last leaf node in the LeafList. The search key is the first one
in the node. However, unlike lookup in a regular trie, when matching of the search key with
an anchor key fails before a leaf node is reached, there is still a chance that the key is
in the index. This is because the keys are stored only at the LeafList and are not directly
indexed by the trie structure. One example is to look up ``Denice'' in the index,
where matching of the first token `D' fails, though the search key is in a leaf node.
Furthermore, when a search key is matched with a prefix of an anchor key,
there is still a chance the search key is not in the index.
An example is to look up ``A'' in the index.

To address the issue, we introduce the concept of \textit{target node} for a search key $K$.
A target node for $K$ is such a leaf node whose anchor key $ K_1$ and immediately next anchor key
$K_2$ satisfy $K_1\leq K<K_2$, if the anchor key $K_2$ exists. Otherwise,
the last leaf node on the LeafList is the search-key's target node. If a search key is in the index,
it must be in its target node. The target nodes of ``A'', ``Denice'', and ``Joseph'' are the first,
second, and fourth leaf nodes in Figure~\ref{fig:arch2}, respectively.
The question is how to identify the target node for a search key.

Looking for a search-key's target node is a process of finding its longest prefix matching an
anchor's prefix. A (short) prefix of the search key can be a prefix of multiple anchors. However,
if its (long) prefix is found to be equal to a unique anchor key, the prefix cannot be another anchor's
prefix due to the prefix condition for being an anchor. Apparently this unique anchor key is not
larger than the search key. Furthermore, if the anchor's next anchor exists, according to anchor's
definition this anchor is smaller than its next anchor and it is not its prefix. However, this
anchor is the search-key's prefix. Therefore, the search key is smaller than the next anchor.
Accordingly, the anchor's leaf node is the target node of the search key. In the example,
the unique anchor of search key ``Joseph'' is ``Jos'',
which can be found by walking down the MetaTrie with the search key.

If there is not such an anchor that is the prefix of a search key, such as ``Denice'' in
Figure~\ref{fig:arch2}, we cannot reach a leaf node by matching token string of the key with
anchor(s) one token at a time starting at the first token. The matching process breaks in
one of two situations. The first one is that a token in the key is found to be non-existent
at the corresponding level of the trie. For example, there isn't an internal node `D' at Level 1
(beneath the root at Level 0) of the trie to match the first token of the search key ``Denice''.
The second one is that tokens of the search key run out during the matching before a leaf node
is reached. An example is with the search key ``A''.

\begin{figure}[!t]
  \centering
  \includegraphics[width=0.8\columnwidth]{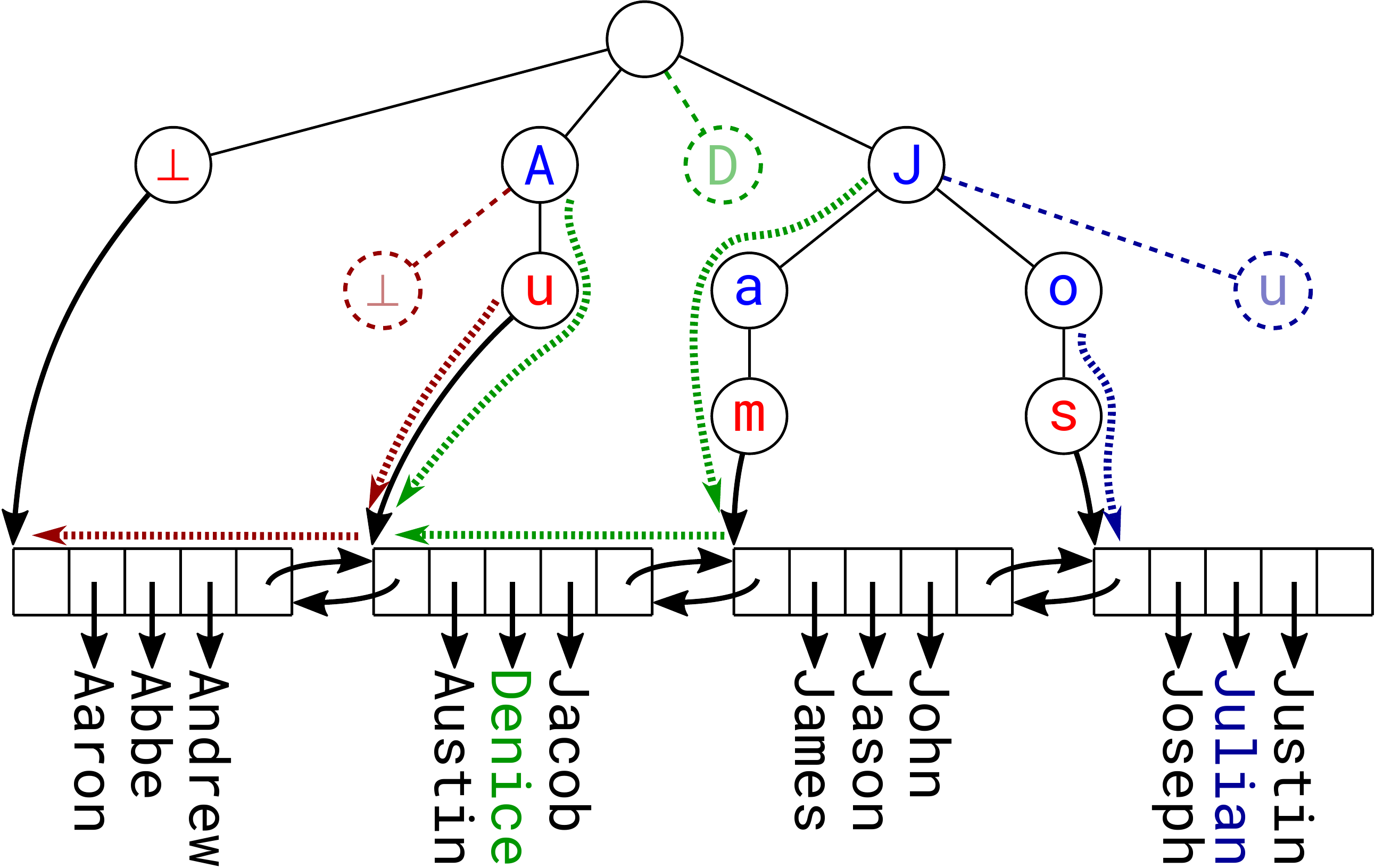}
  \caption{Example lookups on a \textit{MetaTrie} with search keys ``A'', ``Denice'', and ``Julian''.}
  \label{fig:arch2-lookup}
\end{figure}

For the first situation, we assume that a search key's first $k$
tokens ($\langle T_1T_2...T_k\rangle$) are matched and $T_{k+1}$ at Level $k+1$ of
the trie is the first unmatched token. Because $\langle T_1T_2...T_k\rangle$ is not an anchor,
there must exist a node matching
$\langle T_1T_2...T_kL\rangle$, a node matching $\langle T_1T_2...T_kR\rangle$, or both,
where tokens $L < T_{k+1} < R$. In other words, the two nodes are siblings of the hypothetical node
matching $\langle T_1T_2...T_{k+1}\rangle$. Accordingly these two nodes are its left and right siblings,
respectively. We further assume that they are immediate left and right siblings, respectively.
Rooted at left or right sibling nodes there is a subtree, named left or right subtrees, respectively.
If the left sibling exists, the search key's target node is the right-most leaf node of the left subtree.
If the right sibling exists,
the left-most leaf node of the right subtree is the target node's immediate next node on the LeafList.
As all leaf nodes are doubly linked,
the target node can be reached by walking backward on the LeafList by one node.
For search key ``Denice'' in the example, both subtrees exist, which are rooted at internal
nodes ``A'' and ``J'', respectively, and the target node (the second leaf node) can be reached
by either of the two search paths, as depicted in Figure~\ref{fig:arch2-lookup}.
For search key ``Julian'', only the left subtree (rooted at internal node ``O'') is available and only
one search path down to the right-most leaf node exists to reach the target node (the fourth leaf node).

For the second situation, we can append the smallest token $\bot$ on the search key. As we assume
the token is not used in the regular key, $\bot$ becomes the first unmatched key and we can follow
the procedure described for the first situation to find the search-key's target node. Note that in
this case only the right subtree exists. Figure~\ref{fig:arch2-lookup} shows the path to reach
the target node of the search key ``A'', which is the first leaf node.

Once a target node for a search key is identified, further actions for lookup,
insertion, and deletion operations are straightforward.
For lookup, we will search in the target node for the key.
In Section~\ref{sec:wh-opt} we will present an optimization technique to accelerate the process.
Similar to those in the B+ tree, insertion and deletion of a key may lead to splitting
of a leaf node and merging of adjacent leaf nodes to ensure that a node does not grow
over its predetermined capacity and does not shrink below a minimum size.
The difference is that the splitting and merging operations are not (recursively) propagated onto the
parent nodes in Wormhole, as it does in B+ tree, to balance leaf node heights.
The only operations in the MetaTrie are removing anchors for merged nodes
or adding new anchors for split nodes.
To remove an anchor, only the trie nodes exclusively used by the anchor are to be removed.

This composite index structure is more space efficient than a conventional trie index by
storing multiple keys in a leaf node and inserting anchors usually (much) shorter than keys
in the trie. Its search time is practically independent of number of keys in the index,
and is only proportional to anchor lengths, which can be further reduced by intelligently choose
the location where a leaf node is split (we leave this optimization as future work).
However, in the worst case the search time can still be $O(L)$, where $L$ is the length of
a search key. With a long key, the search time can still be substantial. In the following
we will present a technique to further reduce the search cost to $O(\log L)$.

\subsection{Accelerating Search with a Hash Table}
\label{sec:wh-boost}

In the walk from the root of a MetaTrie to a search-key's target leaf node, there are two phases.
The first one is actually to conduct the longest prefix match (LPM) between the search key
and the anchors in the trie. If the longest prefix is not equal to an anchor, the second phase
is to walk on a subtree rooted at a sibling of the token next to the matched prefix of the search key.
The $O(L)$ cost of each of the phases can be significantly reduced.

For the first phase, to obtain the LPM we do not have to walk on the trie along a path from the root
token by token. Waldvogel et al. proposed to use \textit{binary search on prefix lengths}
to accelerate the match for routing table lookups~\cite{WVT97}.
To apply the approach, we insert all prefixes of each anchor into a hash table.
In Figure~\ref{fig:arch2-lookup}'s index, ``Jam'' is an anchor, and accordingly its prefixes
(``'', ``J'', ``Ja'', ``Jam'') are inserted in the hash table. We also track the MetaTrie's height,
or the length of the longest anchor key, denoted $L_{\text{anc}}$.
Algorithm~\ref{alg:bsearch} depicts how a binary search for a search key of length
$L_{\text{key}}$ is carried out.
As we can see the longest prefix can be found in
$O(\log (\text{min}(L_{\text{anc}}, L_{\text{key}})))$ time.
In the example index for search key ``James'' it takes two hash-table lookups
(for ``Ja'' and ``Jam'') to find its longest common prefix (``Jam'').

\begin{algorithm}[t]
\begin{algorithmic}[1]
\caption{Binary Search on Prefix Lengths}
\label{alg:bsearch}
\scriptsize

\Function{searchLPM}{search\_key, L$_{\text{anc}}$, L$_{\text{key}}$}
\State m $\gets$ 0;\quad\quad n $\gets$ min(L$_{\text{anc}}$, L$_{\text{key}}$)+1
\While{(m+1) < n}
  \State prefix\_len $\gets$ (m+n)/2
  \If{search\_key[0 : prefix\_len-1] is in the trie}
    \State m $\gets$ prefix\_len
  \Else\quad
    n $\gets$ prefix\_len
  \EndIf
\EndWhile
\State \Return search\_key[0 : m-1]
\EndFunction
\end{algorithmic}
\end{algorithm}

\begin{figure}[t!]
  \centering
  \includegraphics[width=0.9\columnwidth]{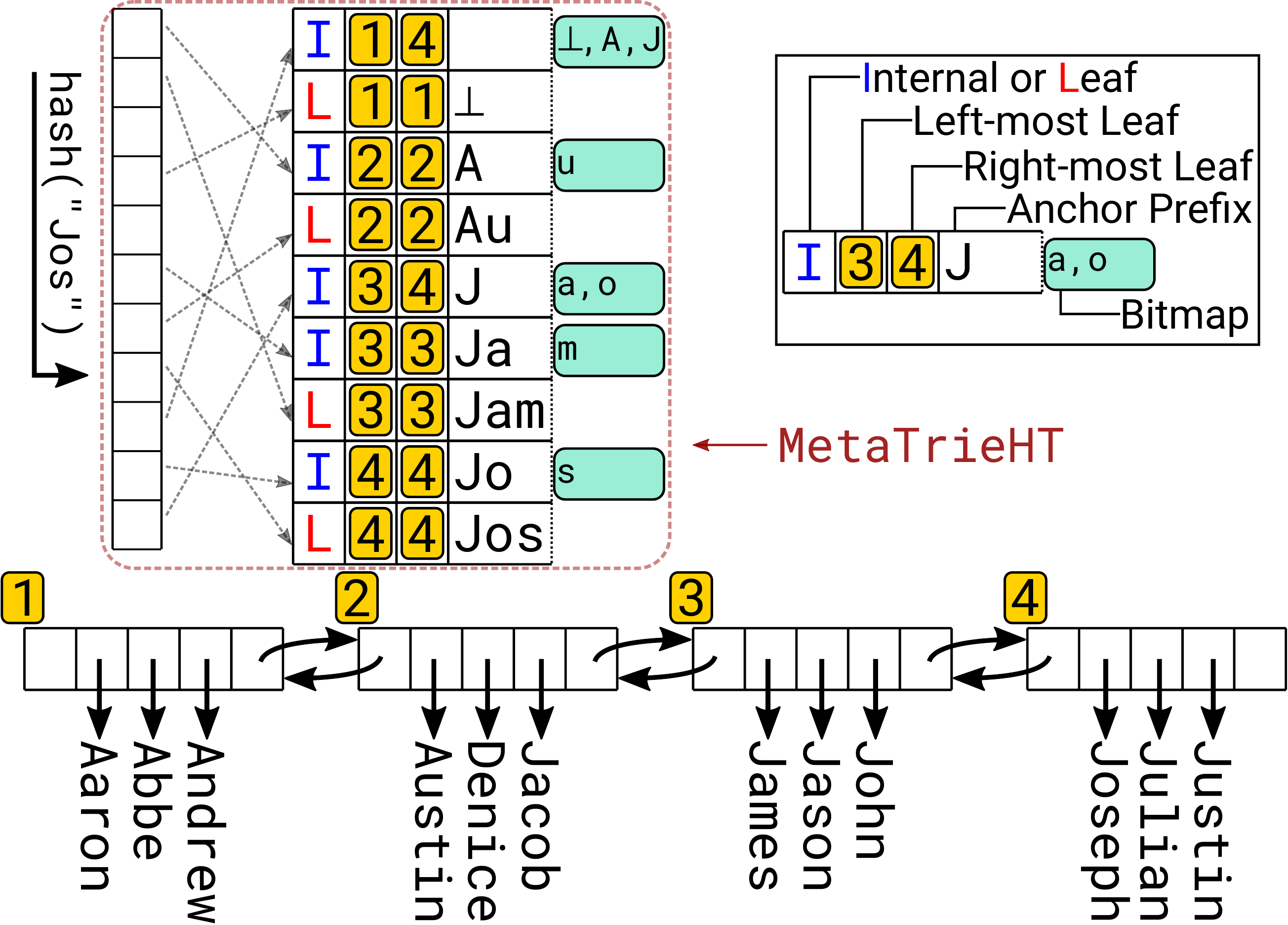}
  \caption{The structure of Wormhole.
For clarity the bitmap is depicted by directly listing child tokens.}
  \label{fig:arch3}
\end{figure}

The hash table is named \textit{MetaTrieHT}, which is to replace the MetaTrie to index
the leaf nodes on the LeafList. The \textit{MetaTrieHT} for the MetaTrie in Figure~\ref{fig:arch2}
is illustrated in Figure~\ref{fig:arch3}. Each node in MetaTrie corresponds to an item in
the hash table. If the node represents an anchor, or a leaf node, the hash item is a leaf item,
denoted `\texttt{L}' in Figure~\ref{fig:arch3}. Otherwise, the node is an internal node and the corresponding hash item is
an internal item, denoted `\texttt{I}'. Using this hash table, pointers in the MetaTrie facilitating
the walk from node to node in the trie are not necessary in the MetaTrieHT,
as every prefix can be hashed into the index structure to know whether it exists.

Each hash item has two fields supporting efficient walk in the second search phase on a path
to a leaf node. The first field is a bitmap. It is meaningful only for internal items.
It has a bit for every possible child of the corresponding internal node in the trie.
The bit is set when the corresponding child exists. With the bitmap, sibling(s) of an unmatched
token can be located in $O(1)$ time. Trie node corresponding to a hash item can be considered
as root of a subtree. In the second phase it is required to know the right-most leaf node
or the left-most leaf node of the subtree. The second field of a hash item contains two pointers,
each pointing to one of the leaf nodes. Accordingly, the second phase takes a constant time.

The index consisting of a LeafList and a MetaTrieHT represents Wormhole's core data structure. Its operations, including lookup (\texttt{GET}), insertion (\texttt{SET}),
deletion (\texttt{DEL}), and range search (\texttt{RangeSearchAscending}) are formally
depicted in Algorithms~\ref{alg:main}, \ref{alg:ancillary}, and~\ref{alg:splitmerge}.
The $O(\log L)$ time cost of Wormhole is asymptotically lower than $O(\log N)$ for B+ tree and
$O(L)$ for trie, where $L$ is the search key's length and $N$ is number of keys in the index.

\begin{algorithm}[t!]
\begin{algorithmic}[1]
\caption{Index Operations}
\label{alg:main}
\scriptsize
\Function{GET}{wh, key}
  \State leaf $\gets$ searchTrieHT(wh, key);\quad\quad
  i $\gets$ pointSearchLeaf(leaf, key)
  \If{(i < leaf.size) and (key = leaf.hashArray[i].key)}
    \State \Return leaf.hashArray[i]
  \Else\ \Return NULL
  \EndIf
\EndFunction

\Function{SET}{wh, key, value}
  \State leaf $\gets$ searchTrieHT(wh, key);\quad\quad
  i $\gets$ pointSearchLeaf(leaf, key)
  \If{(i < leaf.size) and (key = leaf.hashArray[i].key)}
    \State leaf.hashArray[i].value $\gets$ value
  \Else
    \If{leaf.size = MaxLeafSize}
      \State left, right $\gets$ split(wh, leaf)
      \If{key < right.anchor}
        \State leaf $\gets$ left
      \Else\ leaf $\gets$ right
      \EndIf
    \EndIf
    \State leafInsert(leaf, key, value)
  \EndIf
\EndFunction

\Function{DEL}{wh, key}
  \State leaf $\gets$ searchTrieHT(wh, key);\quad\quad
  i $\gets$ pointSearchLeaf(leaf, key)
  \If{(i < leaf.size) and (key = leaf.hashArray[i].key)}
    \State leafDelete(leaf, i)
    \If{(leaf.size + leaf.left.size) < MergeSize}
      \State merge(wh, leaf.left, leaf)
    \ElsIf{(leaf.size + leaf.right.size) < MergeSize}
      \State merge(wh, leaf, leaf.right)
    \EndIf
  \EndIf
\EndFunction

\Function{RangeSearchAscending}{wh, key, count}
  \State leaf $\gets$ searchTrieHT(wh, key);
  \State incSort(leaf.keyArray);\quad\quad out $\gets$ []
  \State i $\gets$ binarySearchGreaterEqual(leaf.keyArray, key)
  \While{(count > 0) and (leaf $\neq$ NULL)}
    \State nr $\gets$ min(leaf.size - i, count);\quad\quad
    count $\gets$ count - nr
    \State out.append(leaf.keyArray[i : (i + nr - 1)])
    \State leaf $\gets$ leaf.right;\quad\quad
    i $\gets$ 0
    \If{leaf $\neq$ NULL} incSort(leaf.keyArray); \EndIf
  \EndWhile
  \State \Return out
\EndFunction

\end{algorithmic}
\end{algorithm}

\begin{algorithm}[t!]
\begin{algorithmic}[1]
\caption{Ancillary Functions}
\label{alg:ancillary}
\scriptsize

\Function{searchTrieHT}{wh, key}
 \State node $\gets$ searchLPM(wh.ht, key, min(key.len, wh.maxLen))
 \If{node.type = LEAF}
   \Return node
 \ElsIf{node.key.len = key.len}
   \State leaf $\gets$ node.leftmost
   \If{key < leaf.anchor} leaf $\gets$ leaf.left
   \EndIf
   \State \Return leaf
 \EndIf
 \State missing $\gets$ key.tokens[node.key.len]
 \State sibling $\gets$ findOneSibling(node.bitmap, missing)
 \State child $\gets$ htGet(wh.ht, concat(node.key, sibling))
 \If{child.type = LEAF}
   \If{sibling > missing} child $\gets$ child.left
   \EndIf
   \State \Return child
 \Else
   \If{sibling > missing}
     \Return child.leftmost.left
   \Else\ \Return child.rightmost
   \EndIf
 \EndIf
\EndFunction

\Function{pointSearchLeaf}{leaf, key}
  \State i $\gets$ key.hash $\times$ leaf.size / (MAXHASH + 1);\quad\quad
  array $\gets$ leaf.hashArray
  \While{(i > 0) and (key.hash $\leq$ array[i - 1].hash)}
    i $\gets$ i - 1
  \EndWhile
  \While{(i < leaf.size) and (key.hash > array[i].hash)}
    i $\gets$ i + 1
  \EndWhile
  \While{(i < leaf.size) and (key.hash = array[i].hash)}
    \If{key = leaf.array[i].key} \Return i \EndIf
    \State i $\gets$ i + 1
  \EndWhile
  \State \Return i
\EndFunction

\Function{allocInternalNode}{initBitID, leftmost, rightmost, key}
  \State node $\gets$ malloc();\quad\quad node.type $\gets$ INTERNAL
  \State node.leftmost $\gets$ leftmost;\quad\quad node.rightmost $\gets$ rightmost
  \State node.key $\gets$ key;\quad\quad node.bitmap[initBitID] $\gets$ 1
  \State \Return node
\EndFunction

\Function{incSort}{array}
  \If{array.sorted.size < THRESHOLD} array $\gets$ qsort(array)
  \Else\ array $\gets$ twoWayMerge(array.sorted, qsort(array.unsorted))
  \EndIf
\EndFunction
\end{algorithmic}
\end{algorithm}

\begin{algorithm}[t!]
\begin{algorithmic}[1]
\caption{Split and Merge Functions}
\label{alg:splitmerge}
\scriptsize

\Function{split}{wh, leaf}
  \State incSort(leaf.keyArray);\quad\quad
  i $\gets$ leaf.size / 2
  \While{\textit{Cannot split between} [i$-1$] \textit{and} [i] \textit{in} leaf.keyArray}
    \State \textit{Try another} i {in range} [1, leaf.size $-1$]
    \State \textit{Abort the split if none can satisfy the criterion}
  \EndWhile
  \State alen $\gets$ commonPrefixSize(leaf.keyArray[i$-1$], leaf.keyArray[i])$+1$
  \State newL $\gets$ malloc();\quad\quad newL.anchor $\gets$ leaf.keyArray[i].key.prefix(alen)
  \State key $\gets$ newL.anchor;\quad\quad \textit{Append 0s to} key {when necessary}
  \State wh.maxLen $\gets$ max(wh.maxLen, key.len)
  \State \textit{Move items at [i to leaf.size$-1$] of leaf.keyArray to newL}
  \State \textit{Insert newL at the right of leaf on the leaf list}
  \State htSet(wh.ht, key, newL)
  \For {plen : 0 to key.len$-1$}
    \State prf $\gets$ key.prefix(plen);\quad\quad
    node $\gets$ htGet(wh.ht, prf)
    \If{node.type = LEAF}
      \State parent $\gets$ allocInternalNode(0, node, node, prf)
      \State htSet(wh.ht, prf, parent);\quad\quad prf.append(0); node.key $\gets$ prf
      \State htSet(wh.ht, prf, node);\quad\quad node $\gets$ parent
    \EndIf
    \If{node = NULL}
      \State node $\gets$ allocInternalNode(key[plen], newL, newL, prf);
      \State htSet(wh.ht, prf, node)
    \Else
      \If{node.leftmost = leaf.right}
        node.leftmost $\gets$ leaf
      \EndIf
      \If{node.rightmost = leaf.left}
        node.rightmost $\gets$ leaf
      \EndIf
    \EndIf
  \EndFor
\EndFunction

\Function{merge}{wh, left, victim}
  \State \textit{Move all items from victim to the left node}
  \State key $\gets$ victim.key;\quad\quad htRemove(wh.ht, key)
  \For {plen : key.len$-1$ to 0}
    \State prefix $\gets$ key.prefix(plen);\quad\quad
    node $\gets$ htGet(wh.ht, prefix)
    \State node.bitmap[key[plen]] $\gets$ 0
    \If{node.bitmap.isEmpty()}
      htRemove(wh.ht, prefix)
    \Else
      \If{node.leftmost = victim}
        node.leftmost $\gets$ victim.right
      \EndIf
      \If{node.rightmost = victim}
        node.rightmost $\gets$ victim.left
      \EndIf
    \EndIf
  \EndFor
\EndFunction
\end{algorithmic}
\end{algorithm}

Regarding space efficiency, Wormhole is (much) better than trie by indexing multiple keys
in a leaf node, rather than individual keys in the trie.
When compared to the B+ tree, it has the same number of leaf nodes.
Therefore, their relative space cost is determined by amount of space held by their respective internal nodes.
Wormhole's MetaTrieHT is essentially organized as a trie,
whose number of nodes highly depends on its key contents.
While it is hard to quantitatively evaluate its space cost and compare it to that of the B+ tree
without assuming a particular workload, we analyze factors impacting the number.
Generally speaking, if the keys often share common prefixes,
many anchors will also share common prefixes,
or nodes on the trie, which reduces the trie size.
On the other hand, if the keys are highly diverse it's less likely to have long common prefixes
between adjacent keys in the LeafList.
According to the rule of forming anchors, short common prefixes lead to short anchors.
Because it is anchors, instead of user keys, that are inserted into the trie,
short anchors lead to fewer internal nodes.
We will quantitatively measure and compare the space costs of Wormhole and B+ tree
with real-world keys in Section~\ref{sec:wh-eval}.

\subsection{Concurrency Support}

To provide strong support of concurrent operations for high scalability, Wormhole aims to minimize
its use of locks, especially big locks, and minimize impact of a lock on concurrency.
There are three groups of operations that require different levels of access exclusiveness.
The first group includes point and range lookups that do not modify the index and do not demand
any access exclusiveness among themselves. The second group includes insertions and deletions
whose required modifications are limited on one or multiple leaf nodes on the LeafList.
They demand access exclusiveness only at the leaf nodes. The third group includes insertions
and deletions that incur split and merge of leaf nodes and modifications of the MetaTrieHT
by adding or removing anchors and their prefixes in it. They demand exclusiveness at the relevant
leaf nodes and at the MetaTrieHT.

A design goal of Wormhole's concurrency control is to minimize the limit imposed by
insertions/deletions on the concurrency of lookup operations.
To this end, we employ two types of locks.
One is a reader-writer lock for each leaf node on the LeafList. For the second group of operations,
insertion/deletion of a key modifies only one leaf node, and accordingly only one node is locked
and becomes unavailable for lookup. For the third group of the operations with one key,
only one or two leaf nodes have to be locked for split or merge, respectively.
However, for addition or removal of prefixes of an anchor in the MetaTrieHT structure,
we may have to simultaneously acquire multiple locks to have exclusive
access of (many) hash items (equivalently trie nodes). To this end the second type of lock
is a single mutex lock on the entire MetaTrieHT to grant exclusive access to an addition or removal
operation of an anchor and its prefixes, instead of fine-grained locks
with much higher complexity and uncertain performance benefits.

However, as every key lookup requires access of the MetaTrieHT table, a big lock imposed
on the entire MetaTrieHT can substantially compromise performance of the first two groups
of operations that perform read-only access on the MetaTrieHT.
To address this issue, we employ the QSBR RCU mechanism~\cite{RCU98,urcu} to enable lock-free access on MetaTrieHT for
its readers (the first two groups of operations).
Accordingly, only the writers of MetaTrieHT need to acquire the mutex lock.
To perform a split/merge operation, a writer first acquires the lock.
It then applies the changes to a second hash table (\textit{T2}),
an identical copy of the current MetaTrieHT (\textit{T1}).
Meanwhile, T1 is still accessed by readers.
Once the changes have been fully applied to T2, T2 will be made visible for readers to access
by atomically updating the pointer to the current MetaTrieHT through RCU,
which simultaneously hides T1 from new readers.
After waiting for an RCU grace period which guarantees T1 is no longer accessed by any readers,
the same set of changes is then safely applied to T1.
Now T1 is again identical to T2 and it will be reused as the second hash table for the next writer.
The extra space used by the second MetaTrieHT is negligible
because a MetaTrieHT, containing only the anchor keys,
is consistently small in size compared with the size of the entire index structure.
As an example, for the eight keysets used in our evaluation (see Table~\ref{tab:datasets}),
the extra space consumed by the second table is only 0.34\% to 3.7\% of the whole index size.

When a lookup reaches a leaf node on the LeafList after searching on a MetaTrieHT,
it needs to make sure that the hash table it used is consistent with the leaf node.
For an insertion/deletion in the third group,
it first acquires lock(s) for relevant leaf node(s), from left to right if two or more leaf nodes are to be locked,
and then acquires the mutex lock for the MetaTrieHT.
With the locks both the leaf node(s) and the table can be updated.
To minimize readers' wait time on the critical section we release the locks
on the leaf nodes right after they have been updated.
To prevent lookups via an old MetaTrieHT from accessing updated leaf nodes,
including nodes being split or deleted, we use version numbers to check their consistency.
Each MetaTrieHT is assigned a version number.
The number is incremented by one for each split/merge operation where a new version of MetaTrieHT is made visible.
Each leaf node is assigned an expected version number, initialized as 0.
When a leaf node is locked for split/merge operation,
we record the current MetaTrieHT's version number plus 1 as the leaf node's expected version number.
A lookup remembers the MetaTrieHT's version number when it starts to access the MetaTrieHT,
and then compares the number with the expected number of the target leaf node it reaches.
If the expected number is greater, this lookup shall abort and start over.

The penalty of the start-overs is limited.
First, for a split/merge operation, only one or two leaf nodes have their version numbers updated.
Lookups targeting any other leaf nodes don't need to start over.
Second, the rate of split/merge is much lower than that of insert/delete operations.
Third, a start-over only needs to perform a second lookup on a newer version of MetaTrieHT,
which is read-only and much faster than an insert/delete operation.

\section{Optimization and Enhancement}
\label{sec:wh-opt}

While Wormhole's design provides significantly improved
asymptotic lookup time, we apply several optimization
techniques to maximize the efficiency of Wormhole's operations on MetaTrieHT
and LeafList. We will also discuss how the assumption on a reserved $\bot$
token not allowed in user keys can be removed. All the techniques
described in this section are also covered in Algorithms~\ref{alg:main},
\ref{alg:ancillary}, and~\ref{alg:splitmerge}.

\subsection{Improving Operations in MetaTrieHT}

There are two major operations in MetaTrieHT for a lookup involving
a sequence of prefixes of a search key. They can be CPU-intensive or
memory-intensive. The first operation is to compute a prefix's
hash value as an index in the MetaTrieHT hash table. The second one
is to read the prefix in the table and compare it with the search-key's
corresponding prefix. Wormhole conducts these operations for each
of its selected prefixes during its binary search for the longest
prefix match. However, a hash-table-based index requires them only
once for a search key. We aim to reduce their CPU and memory access
costs, respectively, and make them comparable with those of the
hash-table-based indexes.

Regarding the first operation, the cost of some commonly used hash
functions, such as that for CRC~\cite{crc} and
xxHash~\cite{xxHash}, is approximately proportional to their
input lengths. By reducing the lengths, we can reduce the hashing
cost. Fortunately, there exist incremental hash functions, including
both CRC and xxHash. Such a function can leverage previously hashed
value of an input string when it computes hash value for an extended
string composed of the string appended with an increment. In this
case it does not need to recompute the longer string from scratch.
Taking advantage of the above properties, Wormhole uses incremental hashing
whenever a prefix match is found and the prefix is extended during
its binary common-prefix search.\footnote{CRC-32c is used in our
implementation.} In this way, the average number of tokens used for
hashing in a lookup of a search key of length $L$ is reduced from
$\frac{L}{2}\log_2 L$ to only $L$, comparable to that of a hash table lookup.

\begin{figure}[t!]
  \centering
  \includegraphics[width=0.9\columnwidth]{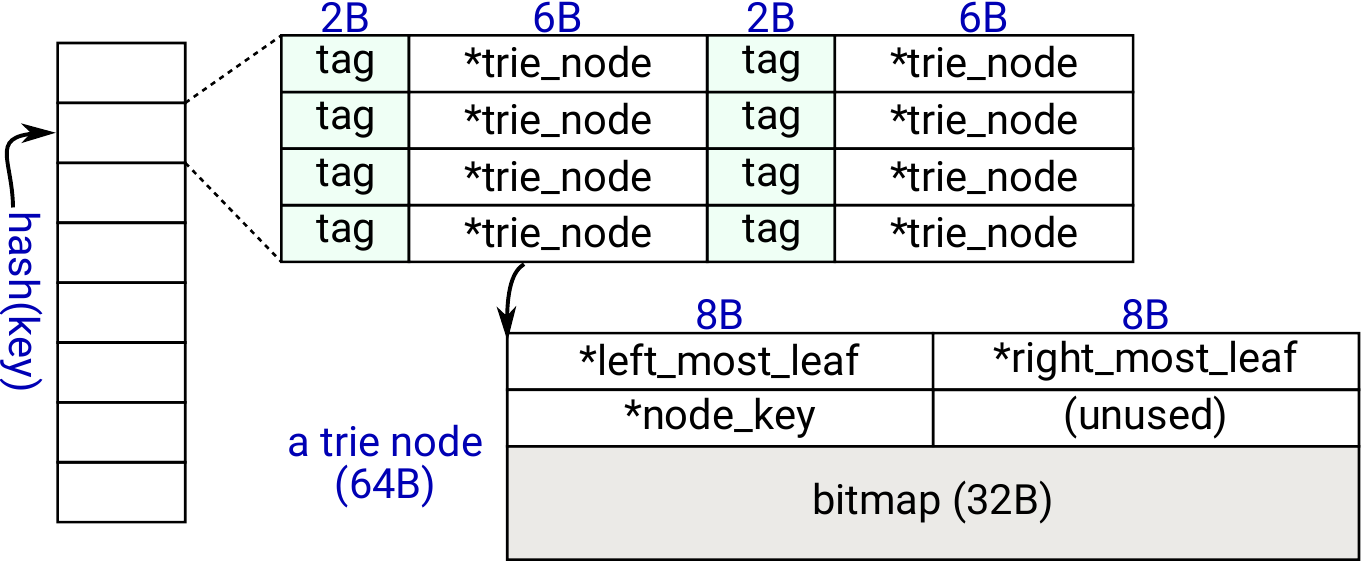}
  \caption{Structure of Wormhole's hash table}
  \label{fig:hash-slot}
\end{figure}

Regarding the second operation, each prefix match operation may
involve multiple prefixes stored in a hash bucket. In the process
many memory accesses may occur,
including dereferencing pointers for prefixes
and accessing potentially long prefixes of several
cache-lines long. These accesses are likely cache misses.
To reduce the cache misses, we organize 8 prefixes in an array
of a cache-line size (64 bytes), named hash slot
(see Figure~\ref{fig:hash-slot}). Each element in the array
consists of a 16-bit tag hashed from the prefix and a 48-bit
pointer to the original prefix.\footnote{On x86-64
only the low-order 48 bits are used in virtual memory address.}
In a lookup, key-comparisons are performed only for prefixes having
a matched tag, which effectively reduces  average number of
key-comparisons to almost one per lookup. Similar approaches have been
widely used in high-performance hash tables~\cite{FAK13,BZG16,LAK14}.

However, it takes multiple hash-table lookups to find an LPM,
which still leads to multiple key-comparisons for a lookup on Wormhole.
To further reduce this overhead,
we first optimistically trust all tag-matches and omit key-comparisons
in every hash-table lookup until finding a seemingly correct LPM.
Tag comparisons may produce false-positive matches, which can
lead the binary search to a wrong prefix that is longer than
the correct one.
To detect this error, a full key comparison is
performed at the last prefix after the binary search.
If it is a mismatch, the search will start over with full prefix comparisons.
Note that there are no false-negative matches in this approach.
Accordingly, it always produces the correct longest prefixes if false-positive matches
do not occur. With the 16-bit tags produced by a well-designed hash
function, the probability of error occurrence is only
$0.0153\%$ for keys of 1024-bytes long ($1-(\frac{2^{16}-1}{2^{16}})^{10}$).

\subsection{Improving Operations in Leaf Node}

Once a target leaf node is identified, a search of a key within
the node is carried out. As keys in the node are sorted, a binary
search may be used during the search. Similar to the issue of many
memory accesses in the MetaTrieHT, accessing a number of original
(long) keys for comparison can be very expensive. Accordingly, we also
calculate a 16-bit hash tag for each key and place the tags in a tag
array in the ascending hash order. A search is then conducted on the
compact tag array. Only when a tag is matched will its corresponding
key be read and compared, which substantially reduces the number of
memory references.

We then further reduce number of comparisons on the tag array using a direct
speculative positioning approach. If a hash function that uniformly
hashes keys into the tag space is employed, the tag values themselves
are well indicative of their positions in the array. Specifically,
with a tag of value $T$ computed from a search key we will first
compare it with a tag at position $\frac{k\times T}{T_{\text{max}}+1}$
in the key array, where $k$ is number of keys in the array and
$T_{\text{max}}$ is the largest possible tag value.
If there isn't a match at the position, we will compare it with its neighboring
tags. Using the lower 16-bits of a (CRC-32c) hash value as the tag,
it usually takes only 1 to 3 tag comparisons
to complete the search in a node of 128 keys.

\begin{figure}[t!]
  \centering
  \includegraphics[width=0.8\columnwidth]{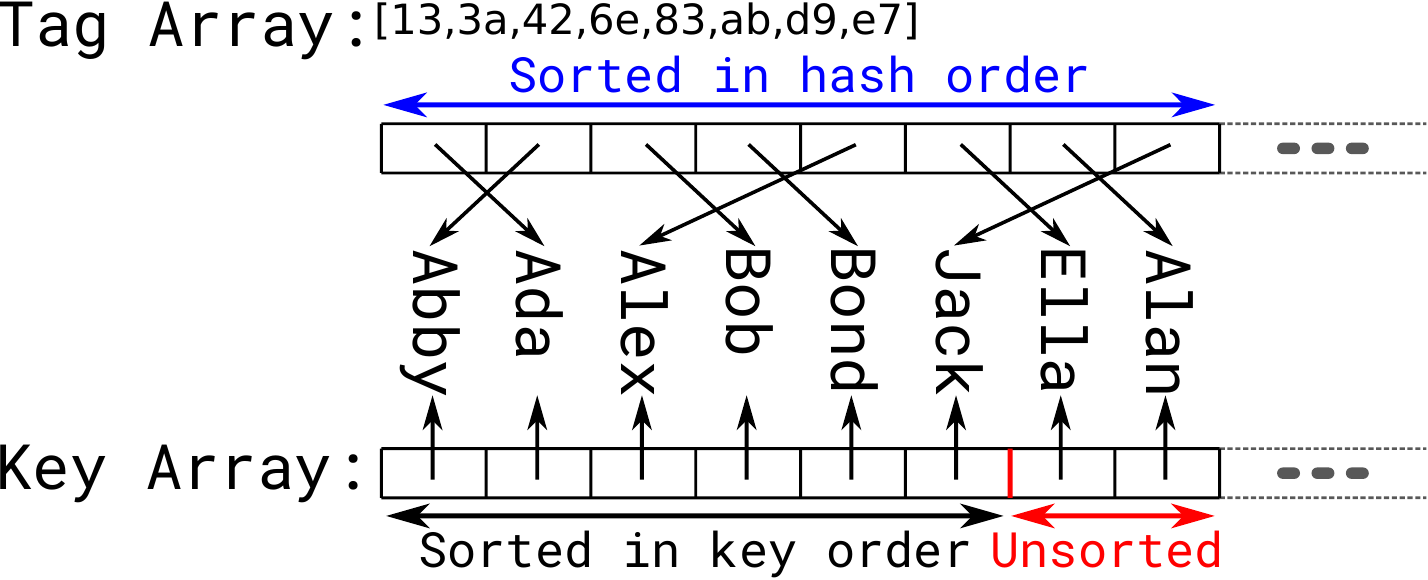}
  \caption{Wormhole's leaf node}
  \label{fig:wh-leaf}
\end{figure}

Another benefit of having the compact tag array
is that the original key array does not have to always stay sorted.
For efficiency, we may append newly inserted keys after the
keys in the key array without immediate sorting,
as illustrated in Figure~\ref{fig:wh-leaf}.
The sorting on the key array can be indefinitely delayed
until a range search or split reaches the node.
Further, the batched sorting amortizes the cost of ordered insertions
when multiple unsorted keys are appended.

\subsection{Wormhole with Any Key Tokens}
\label{sec:zero-keys}

We have assumed existence of a token value that never appears in
regular keys, similar to an assumption in the design of
Masstree~\cite{MKM12}. With this assumption, we had designated
an unused value, denoted `$\bot$', as the smallest value and
used it to extend prefix so as to form an anchor satisfying the
rule that no anchor can be a prefix of another anchor.
By removing the assumption, we have to allow the minimal token value, say binary zero,
to appear in the keys.
This is not an issue for printable keys where 0 is not used.
However, a difficult situation arises for binary keys when a
potential anchor (generated due to node split) becomes a prefix of
another anchor that consists of the prefix and trailing zeroes.

\begin{figure}[t!]
  \centering
  \includegraphics[width=0.45\columnwidth]{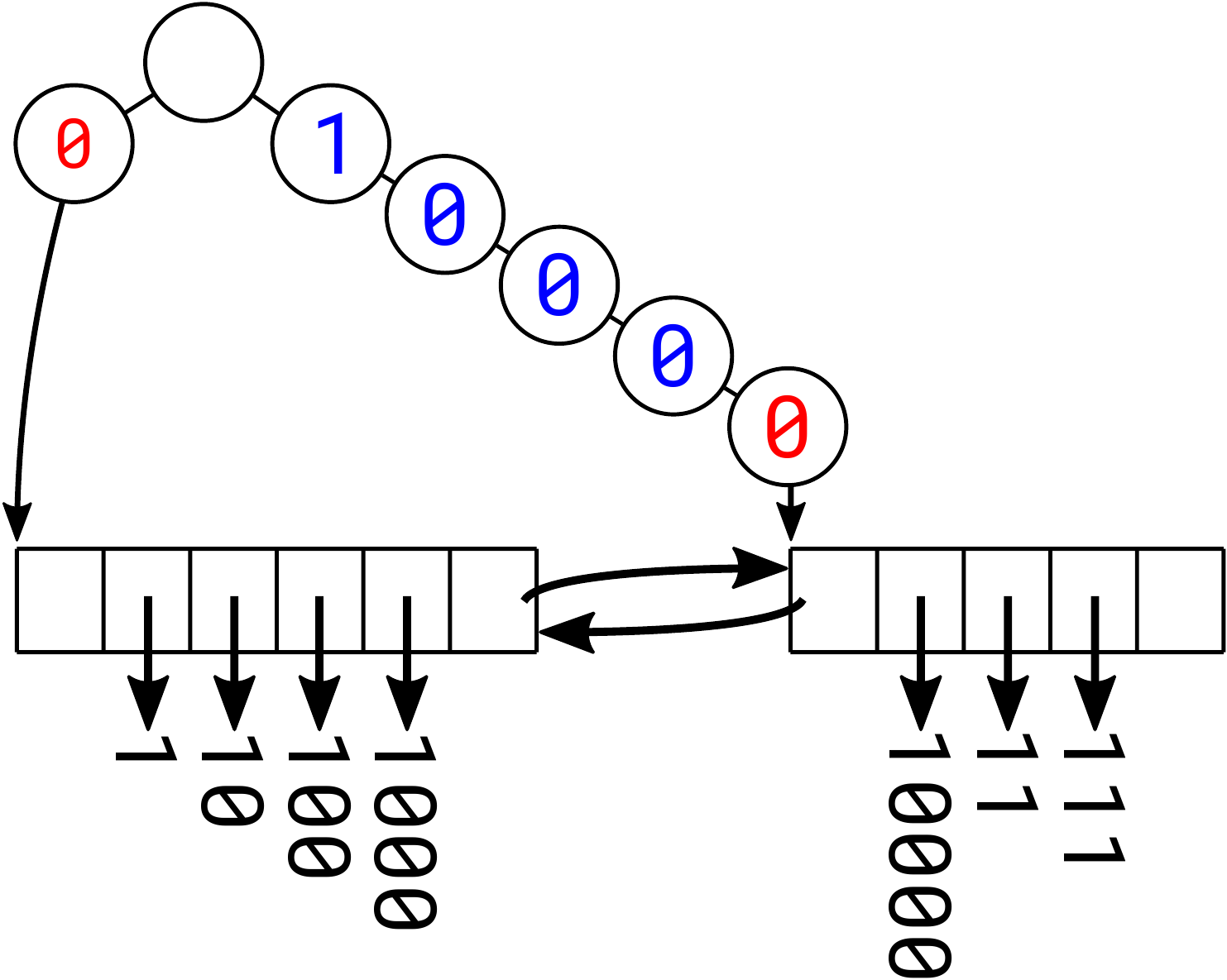}
  \caption{Introducing a fat leaf node}
  \label{fig:zero-keys}
\end{figure}

One example is that we cannot identify any position in the first
leaf node in Figure~\ref{fig:zero-keys} to split it and produce a
legitimate anchor. Suppose we split the node in the middle and
select binary ``100'' as the anchor. Apparently it is a prefix of the next
anchor ``10000'' and it violates the prefix condition. In this case where
all keys in the node are composed of a common prefix ``1'' and a number of trailing `0's,
there is not a position where we can split and form a new anchor.
To address this issue, we simply allow the leaf node to grow over the node
capacity into a \textit{fat node} without splitting it.
Note that the introduction of fat node is mainly for correctness and
we believe it has virtually no impact on real systems.
For example, with a maximal node size of $N$, having a fat node requires that
there are at least $N+1$ keys sharing the same prefix but having different numbers of trailing zeroes.
In this case the longest key among them must have at least $N$ trailing zeroes.
With a moderate $N$ of 64 or 128, the fat node is unlikely to be seen with any real datasets.

\section{Evaluation}
\label{sec:wh-eval}

In this section we experimentally evaluate Wormhole by comparing it with several
commonly used index structures, including B+ tree~\cite{C79}, skip list~\cite{skiplist},
Adaptive Radix Tree (ART)~\cite{LKN13}, and Masstree~\cite{MKM12}.

In the Wormhole prototype we use 128 as the maximum leaf-node size (number of keys in a leaf node).
We use an STX B+-tree~\cite{stxbptree}, a highly optimized in-memory B+ tree implementation, to accommodate large datasets.
The B+ tree's fanout is set to 128, which yields the best result on our testbed.
We use the skip list implementation extracted from LevelDB~\cite{leveldb}.
ART is a trie-like index with a lookup cost of $O(L)$.
To reduce space consumption, ART adaptively selects its node size and
employs path compression to reduce number of nodes.
We use an ART's implementation available on Github~\cite{libart}.
Masstree is a trie-like index with a very high fanout (up to $2^{64}$).
With this high fanout it is impractical to use arrays to hold children pointers in trie nodes.
Therefore, at each trie node it employs a B+ tree to index the children.
We use the publicly available source code of Masstree from
its authors in the evaluation~\cite{masstree}.

Among the five indexes, only Wormhole and Masstree employ fine-grained
RCU and/or locks, which enables thread-safe access for all of their index operations.
The other three indexes are not designed with built-in concurrency control mechanisms.
For example, LevelDB needs to use an external mutex lock to synchronize writers on its skip list.
For fair comparison, we only compare Wormhole with their thread-unsafe implementations
with read-only or single-writer workloads.

Experiments are run on a Dell R630 server with two 16-core Intel Xeon E5-2697A v4 CPUs,
each with 40\,MB LLC. To minimize the interference between threads or cores,
hyper-threading is turned off from BIOS and we use one NUMA node to run the experiments.
The server is equipped with 256\,GB DDR4-2400 ECC memory (32\,GB$\times$8) and
runs a 64-bit Linux (v4.15.15).
To evaluate Wormhole in a networked environment, we connect two identical servers
of the above configuration with a 100\,Gb/s Infiniband (Mellanox ConnectX-4).
Requests of index operations are generated from one server and are sent to the other for processing.

\begin{table}[t!]
\centering
\footnotesize
\caption{Description of Keysets} \label{tab:datasets}
\begin{tabular}{l|l|r|r}
\toprule
\makecell[c]{Name} & \makecell[c]{Description} & \makecell[c]{Keys \\
  ($\times 10^6$)} & \makecell[c]{Size \\ (GB)} \\
\hline
Az1  & \makecell[l]{Amazon reviews metadata, \\
  avg. length: 40\,B, format: \textit{item-user-time}} & 142 & 8.5 \\
\hline
Az2  & \makecell[l]{Amazon reviews metadata; \\
  avg. length: 40\,B, format: \textit{user-item-time}} & 142 & 8.5 \\
\hline
Url & \makecell[l]{URLs in Memetracker,
  avg. length: 82\,B} & 192 & 20.0 \\
\hline
K3   & Random keys, length: 8\,B & 500 & 11.2 \\
\hline
K4   & Random keys, length: 16\,B & 300 & 8.9 \\
\hline
K6   & Random keys, length: 64\,B & 120 & 8.9 \\
\hline
K8   & Random keys, length: 256\,B & 40 & 10.1 \\
\hline
K10  & Random keys, length: 1024\,B & 10 & 9.7 \\
\bottomrule
\end{tabular}
\end{table}

We use publicly available datasets collected at Amazon.com~\cite{MY16} and MemeTracker.org~\cite{meme9}.
The original Amazon dataset contains 142.8 million product reviews with metadata,
We extract three fields (Item ID, User ID, and Review time) in the metadata to construct two keysets,
named \textit{Az1} and \textit{Az2}, by concatenating them in different orders
(see Table~\ref{tab:datasets}).  Key composition varies with the order,
and may impact the index's performance, especially for the trie-based indexes
(B+ tree and Wormhole). For the MemeTracker dataset we extract URLs from it and
use them as keys in the keyset, named \textit{Url}.

For trie-based indexes a performance-critical factor is key length. We create five synthetic keysets,
each with a different fixed key length (from 8\,B to 1024\,B). Key count is selected to make sure
each keyset is of the same size (see Table~\ref{tab:datasets}). Key contents are randomly generated.

In the evaluation we are only concerned with performance of index access and skip access of
values in the KV items. In the experiments, the search keys are uniformly selected
from a keyset to generate a large working set so that an index's performance is not overshadowed
by effect of CPU cache. In the experiments we use 16 threads to concurrently access the indexes
unless otherwise noted.

\subsection{Lookup Performance}
\label{sec:eval-lookup}

In the experiments for measuring lookup throughput we insert each of the keysets to an index,
then perform lookups on random keys in the index.

\begin{figure}[!t]
  \centering
  \includegraphics[width=0.9\columnwidth]{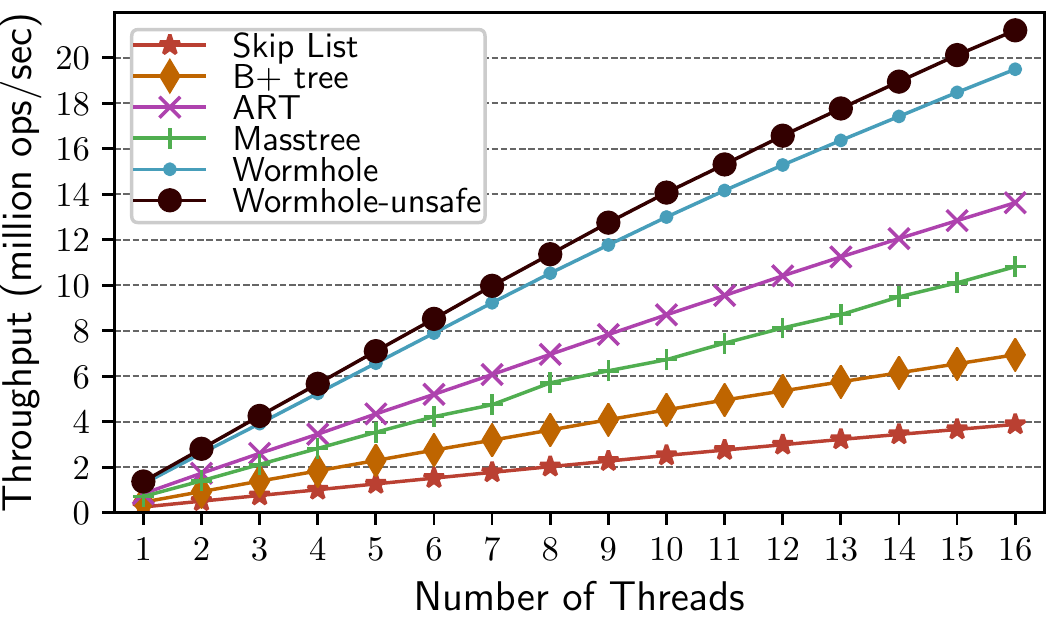}
  \caption{Lookup throughput with different number of threads.
  The \textit{Az1} keyset is used in this experiment.}
  \label{fig:threads}
\end{figure}

We first measure single-thread throughput of the indexes
and see how they scale with number of the threads.
The results with \textit{Az1} keyset are shown in Figure~\ref{fig:threads}.
With one thread, Wormhole's throughput is 1.266\,MOPS (million operations per second),
about 52\% higher than that of ART (0.834\,MOPS), the second-fastest index in this experiment.
All of the five indexes exhibit good scalability.
As an example, Wormhole's throughput with 16 threads (19.5\,MOPS) is 15.4$\times$ of that with one thread.
In addition, it's 43\% higher than that of ART with 16 threads.
We also create a thread-unsafe version of wormhole index (namely \textit{Wormhole-unsafe})
by not using of the RCU and the locks.
As shown in Figure~\ref{fig:threads},
the thread-unsafe Wormhole reaches 21.2\,MOPS, a 7.8\% increase of its thread-safe counterpart.
Since the results of the other keysets all show a consistent trend as described above,
we omit them from this paper.

\begin{figure}[!t]
  \centering
  \includegraphics[width=0.9\columnwidth]{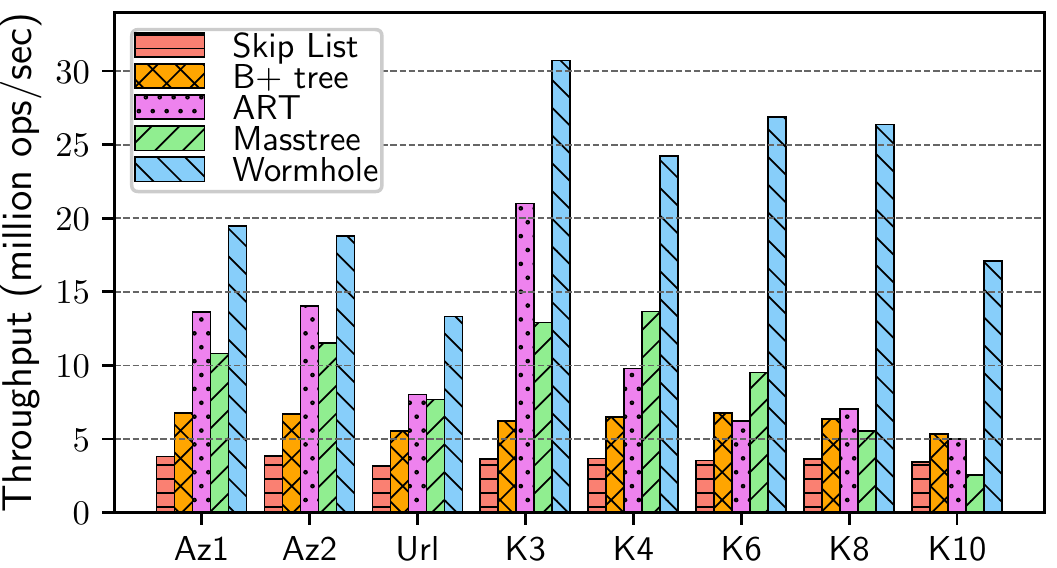}
  \caption{Lookup throughput on local CPU}
  \label{fig:tpyh16}
\end{figure}

We then investigate Wormhole's performance with different keysets.
We use 16 threads for the following experiments unless otherwise noted.
The throughput results with the eight keysets are shown in Figure~\ref{fig:tpyh16}.
Wormhole improves the lookup throughput by 1.3$\times$ to 4.2$\times$ when compared
with the best results among the other indexes for each keyset.
Compared with throughput of the B+ tree and skip list, throughput of the other three indexes exhibits higher
variations due to their use of trie structure and variable key lengths in different keysets.
In the meantime, throughput of Masstree and ART are more tightly correlated to key length than Wormhole.
Masstree substantially outperforms B+ tree and skip list with short keys
(e.g., \textit{K3} and \textit{K4}). However,
its throughput drops quickly with longer keys (e.g., \textit{Url}, \textit{K8}, and \textit{K10}).
Wormhole's lookup throughput is much less affected by key-length variation because
its throughput is determined by the anchor length ($L_{\text{anc}}$), which is usually (much) smaller
than the average key length. Specifically, it is determined by 
$\log (\text{min}(L_{\text{anc}}, L_{\text{key}}))$, rather than by $L$ in Masstree and ART
(see Algorithm~\ref{alg:bsearch}).
In \textit{Url} the URLs often share long common prefixes, which leads to long anchors
(about 40\,B in average as measured) in Wormhole.
Even though, Wormhole still outperforms
the others by at least 1.7$\times$.

\begin{figure}[!t]
  \centering
  \includegraphics[width=0.9\columnwidth]{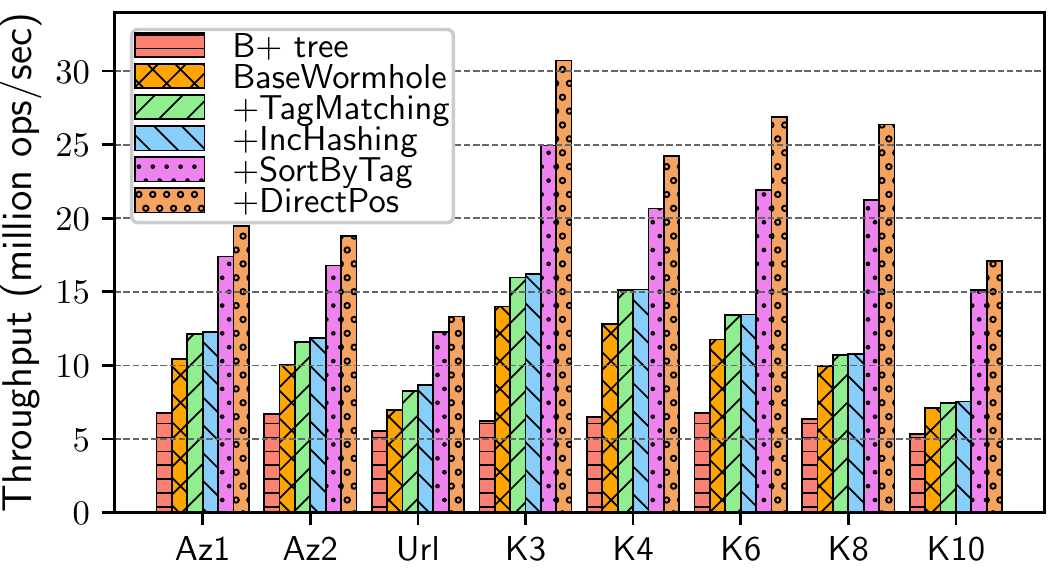}
  \caption{Throughput with optimizations applied. For an optimization,
    the ones above it in the legends are also applied.
    E.g., \texttt{+DirectPos} represents all optimizations are applied.}
  \label{fig:wh-break}
\end{figure}

Various optimizations are applied in Wormhole's implementation,
including tag matching in MetaTrieHT (\textit{TagMatching}), incremental hashing (\textit{IncHashing}),
sorting by tags at leaf nodes (\textit{SortByTag}), and direct speculative positioning in the leaf nodes (\textit{DirectPos}).
To see how much individual optimizations quantitatively contribute to the Wormhole's improvement,
we incrementally apply them one at a time to a basic Wormhole version without the optimizations
(\textit{BaseWormhole}). Figure~\ref{fig:wh-break} shows the throughput of Wormholes without and with
the incrementally added optimizations, as well as that of B+ tree as a baseline on different keysets.
As shown, BaseWormhole improves the throughput by 1.26$\times$ to 2.25$\times$.
After two optimizations (TagMatching and IncHashing) are applied,
the improvement increases to 1.4$\times$ to 2.6$\times$. The index workloads are memory-intensive,
and memory access efficiency plays a larger role than CPU in an index's overall performance.
As TagMatching reduces memory accesses, and corresponding cache misses,
it contributes more to throughput improvement than IncHashing,
which reduces CPU cycles and has a contribution of only about 3\%.
A more significant improvement is received with SortByTag and DirectPos applied at the leaf nodes.
At the leaf nodes SortByTag removes expensive full key comparisons. Its contribution is bigger
with keysets of longer keys. DirectPos can dramatically reduce number of tag comparisons from 6--7 to less than 3 (on average), and also substantially contributes to the throughput improvements
(though less significant than SortByTag). Overall with all the optimizations the throughput
is improved by up to 4.9$\times$ by Wormhole.

\begin{figure}[!t]
  \centering
  \includegraphics[width=0.9\columnwidth]{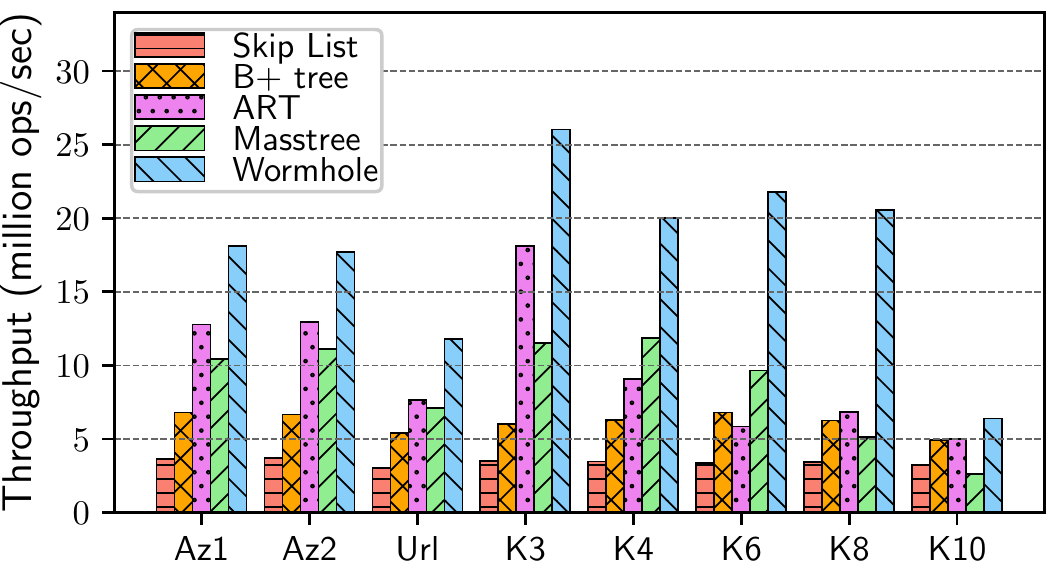}
  \caption{Lookup throughput on a networked key-value store}
  \label{fig:tpib16}
\end{figure}

Network had often been considered as a major potential bottleneck for client/server applications and
a slow connection can overshadow any performance improvement made at the host side.
However, today's off-the-shelf network devices are able to offer a high bandwidth
close to the speed of main memory.
For example, the aggregated bandwidth of three 200\,Gb/s Infiniband (IB) links
(3$\times$24\,GB/s) is close to that of a CPU's memory controller
(76.8GB/s for a Xeon E4 v4 CPU).
This ever-increasing network bandwidth makes performance of networked applications
more sensitive to the efficiency of the host-side CPU/memory usage.
To evaluate by how much Wormhole can improve performance of networked data-intensive applications,
we port the indexes to HERD, a highly optimized RDMA-enabled key-value store~\cite{rdmabench},
and run the lookup benchmarks over a 100\,Gb/s IB link.
We use a batch size of 800 (requests per operation) for RDMA sends and receives.
The throughput results are shown in Figure~\ref{fig:tpib16}.
Generally speaking, Wormhole is able to maintain its advantage over the other indexes,
which is comparable to the results on a single machine (Figure~\ref{fig:tpyh16}).
However, the peak throughput of Wormhole is decreased by 5\% to 20\% for most datasets.
For the \textit{K10} dataset, the large key size (1\,KB each) significantly inflates the size of each request.
In this setting with one IB link, the network bandwidth becomes the bottleneck that
limits the improvement of Wormhole.
As a result, with the \textit{K10} dataset Wormhole's throughput is only 37.5\% of that without the network,
and is only 30\% higher than that of B+ tree.

\subsection{Comparing with Hash Tables}

\begin{figure}[!t]
  \centering
  \includegraphics[width=0.65\columnwidth]{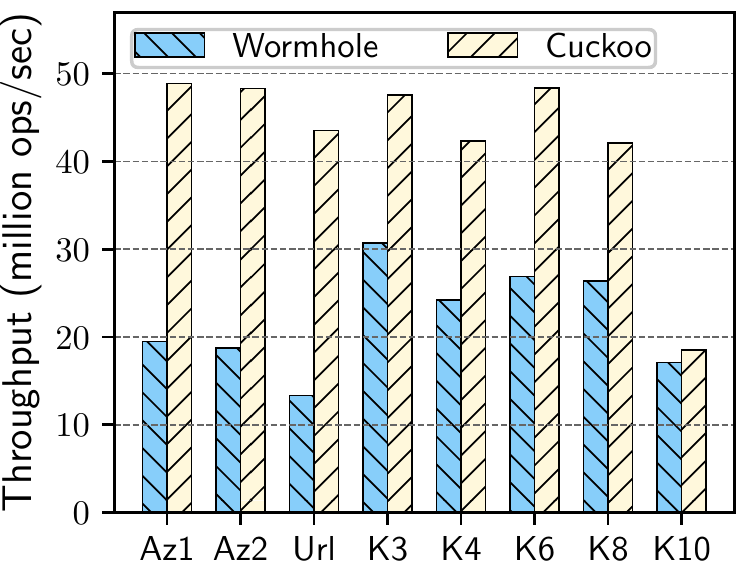}
  \caption{Lookup throughput of Wormhole and Cuckoo hash table}
  \label{fig:wh-cuckoo}
\end{figure}

Wormhole aims to bridge the performance gap between ordered indexes and hash tables.
To know how far Wormhole's performance is close to that of hash tables,
we compare Wormhole with a highly optimized Cuckoo hash table~\cite{libcuckoo}.
The experimental results are shown in Figure~\ref{fig:wh-cuckoo}.
For the first seven keysets, Wormhole's throughput is about 31\% to 67\% of that of the hash table.
The $K10$ keyset has very long keys (1024-byte keys).
16 cache-lines need to be accessed in one key comparison. And the key-access cost dominates
lookup time in both indexes. By using only tags in the MetaTrieHT and leaf nodes in the comparison
in both Wormhole and the optimized hash table, they have similar number of full key accesses.
As a result, on this keyset Wormhole's throughput is close to that of the hash table.

\begin{figure}[!t]
  \centering
  \includegraphics[width=0.9\columnwidth]{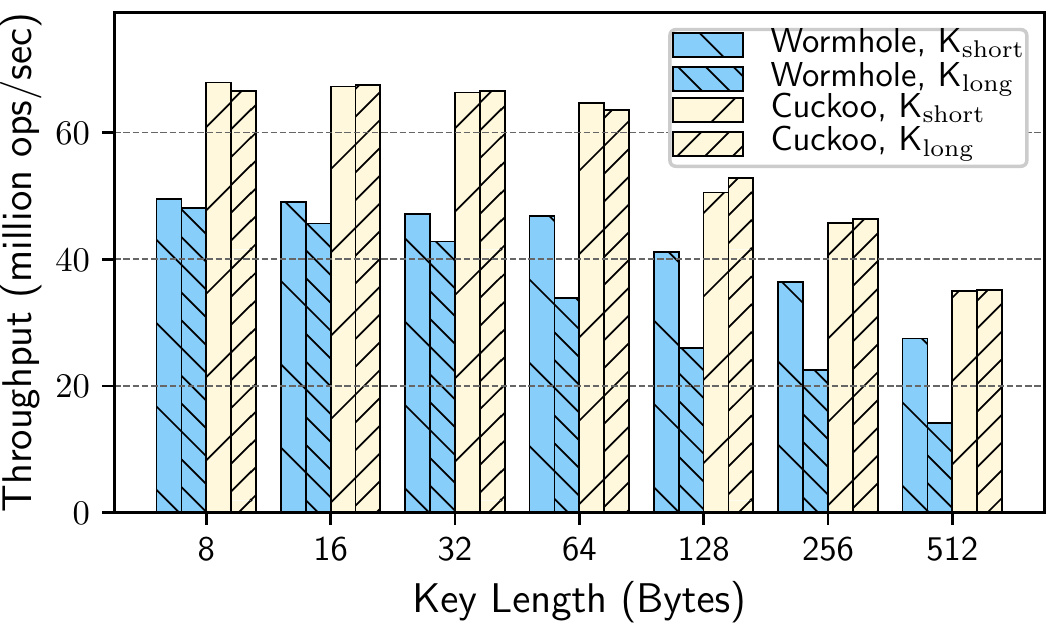}
  \caption{Lookup throughput for keysets of short and long common prefixes}
  \label{fig:wh-lohi}
\end{figure}

Besides key length, another factor affecting Wormhole's lookup efficiency is anchor length,
which determines the MetaTrieHT's size and lookup time in MetaTrieHT. With randomly generated
key contents, the anchors are likely very short. However, in reality a key's true content may
only occupy the last several bytes and fill the leading bytes of a key with the same filler token
such as `0'. To simulate this scenario, we form a number of keysets. Each keyset contains
10 million keys of a fixed size ($L$). Such a keyset, denoted as  ($K_{\text{short}}$),
contains keys of random contents and is expected to have short anchors.
We then fill each-key's first $L-4$ bytes with `0', and denote the resulting keyset as $K_{\text{long}}$.
Figure~\ref{fig:wh-lohi} shows lookup throughput on the two keysets, $K_{\text{short}}$ and $K_{\text{long}}$,
at different key lengths.

The Cuckoo hash table shows little throughput difference with the two keysets at various key lengths.
However, with longer anchors Wormhole's throughput on $K_{\text{long}}$ is lower than that on $K_{\text{short}}$.
This throughput reduction becomes larger with long keys. With the longest keys (512\,B)
the corresponding long anchors lead to more memory accesses
(e.g., $\log_2 512 = 9$ for LPM on the MetaTrieHT),
reducing its throughput from about 78\% of the hash-table's throughput to only 40\%.

\subsection{Performance of other Operations}
\label{sec:other-ops}

In this section we use workloads having insertion operations.
Note that several indexes we used (skip list, B+ tree, and ART) cannot safely perform concurrent
accesses when a writer is present.
If we apply locking or use their lockfree/lockless variants to allow concurrent readers and writers,
their performance can be penalized because of the extra overhead.
For a more vigorous and fair comparison,
we compare Wormhole with their implementations without concurrency control.
Accordingly, we use only one thread for insertion-only workloads,
and then exclude the three thread-unsafe indexes in the evaluation with multi-threaded read-write workloads.

\begin{figure}[!t]
  \centering
  \includegraphics[width=0.9\columnwidth]{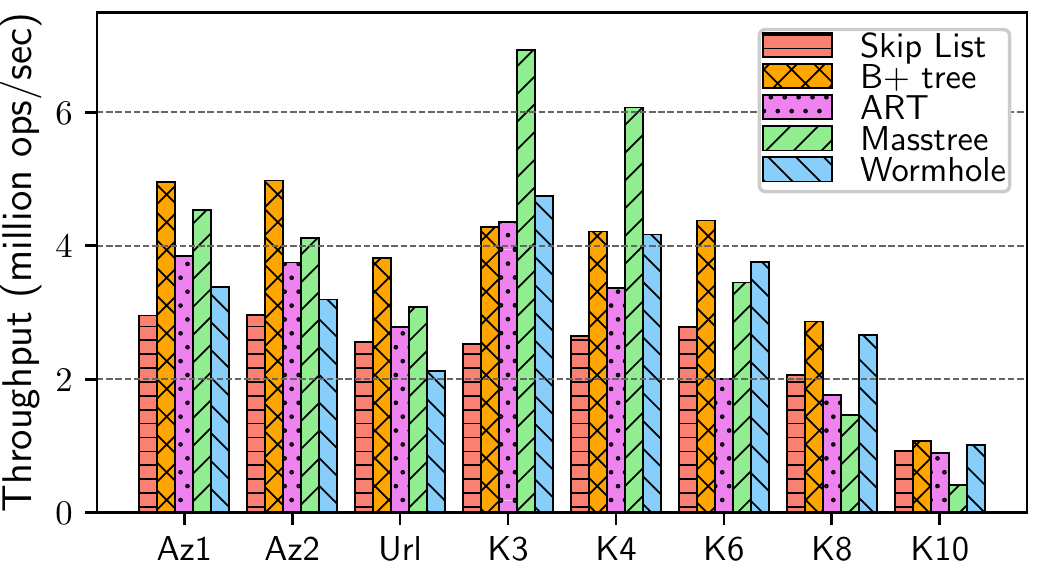}
  \caption{Throughput of continuous insertions}
  \label{fig:wh-insert}
\end{figure}

In insertions-only workloads, keys from
a keyset are inserted into an initially empty index, and the insertion throughput is shown in Figure~\ref{fig:wh-insert}. Wormhole's throughput is comparable
to that of the skip list on most keysets.
With short keys (e.g., \textit{K3} and \textit{K4}),
both Masstree and Wormhole show a higher throughput than comparison-based indexes
(B+ tree and skip list) as insertion of short keys has a low cost on a trie-like structure.
However, with longer keys (e.g., \textit{Url}) throughput of Masstree and Wormhole becomes lower.

\begin{figure}[!t]
  \centering
  \includegraphics[width=0.9\columnwidth]{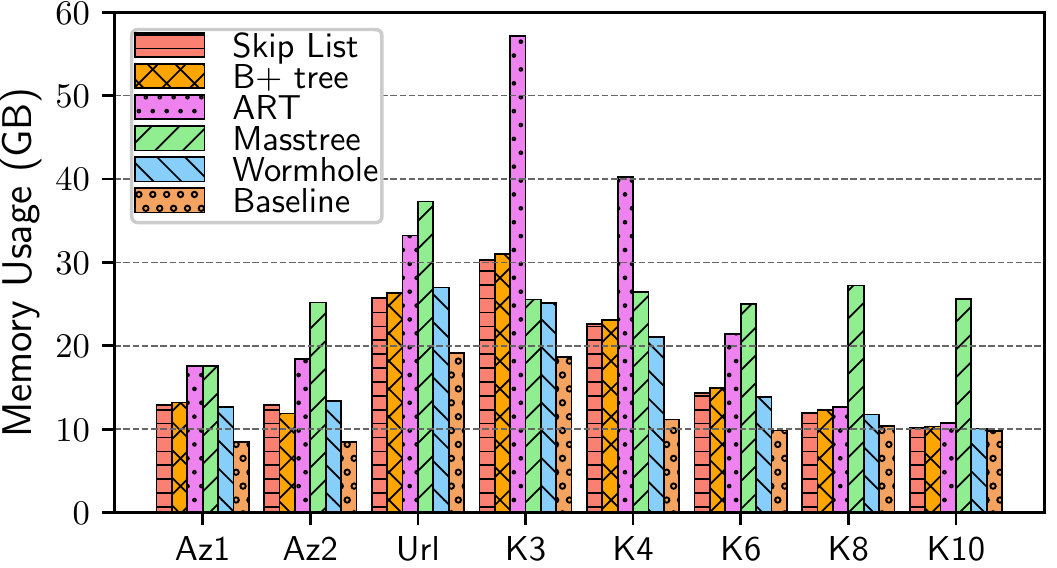}
  \caption{Memory usage of the indexes}
  \label{fig:wh-size}
\end{figure}

When an index for each of the keysets is built, we estimate their memory demands by
taking difference of resident memory sizes, reported by the \texttt{\small getrusage()}
system call, before and after an index is built.
Hugepages are disabled for this experiment to minimize memory wastage due to internal fragmentation.
In the indexes, space for each KV item is allocated separately and is reached with a pointer in an index node.
To establish a \textit{baseline} to represent minimal memory demand of a keyset,
we multiply the key count of the set with the sum of key length and a pointer's size.
Memory demands of the indexes are shown in Figure~\ref{fig:wh-size}.
As shown, in most cases Wormhole's memory usage is comparable to
those of B+ tree and skip list.
Wormhole uses a small trie to organize its anchors and places the keys in
large leaf nodes. As anchors can be much shorter than keys, the space overhead
of the MetaTrieHT can be further reduced, leading to a higher space efficiency than the trie-based Masstree,
which places keys in the trie structure.
Masstree's memory usage is significantly higher than the other indexes, except for
keysets with very short keys (e.g., \textit{K3}) where the entire index is actually managed
by a single B+ tree at the root trie node.
On the contrary, ART has significantly higher space consumption with short keys
(\textit{K3} and \textit{K4}), which is due to its excessive number of trie nodes.
With longer keys, the path compression helps to amortize the space cost
with relatively reduced numbers of trie nodes.

\begin{figure}[!t]
  \centering
  \includegraphics[width=0.9\columnwidth]{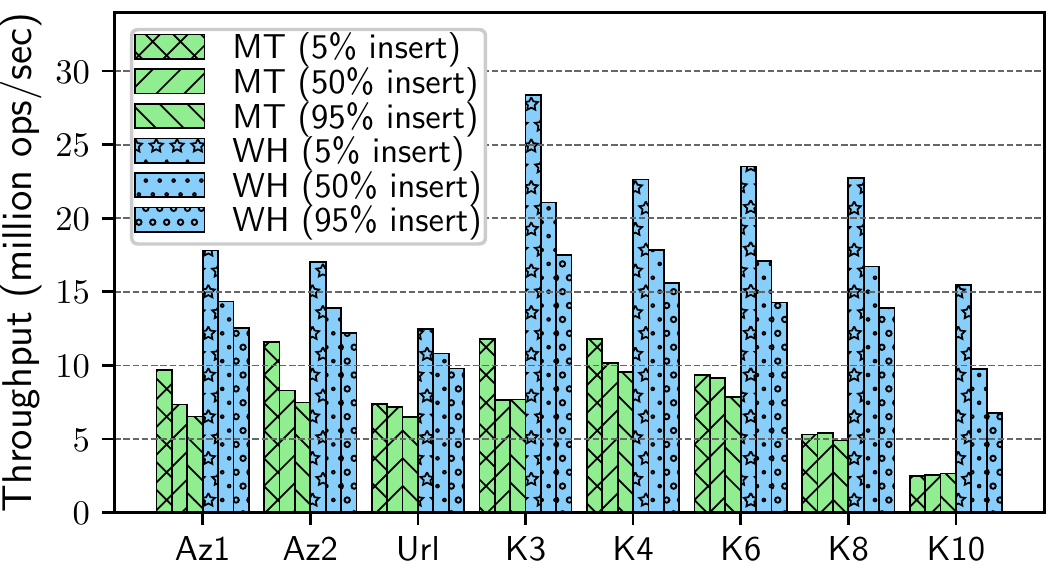}
  \caption{Throughput of mixed lookups and insertions}
  \label{fig:wh-rw}
\end{figure}

We now evaluate Wormhole with workloads of mixed lookups and insertions using 16 threads.
As shown in Figure~\ref{fig:wh-rw}, we change percentage of insertions from 5\%, 50\%, to 95\% of the total operations to see
how Wormhole's performance is affected by operations that may update the MetaTrieHT.
In general, the trend of relative throughput between Masstree and Wormhole with insertions
on different keysets is similar (compare Figures~\ref{fig:tpyh16} and~\ref{fig:wh-rw}).
With more insertions, the throughput improvements of Wormhole over Masstree become smaller,
but still substantial. With a big leaf node most insertions do not update the MetaTrieHT
and lookup time still holds a significant portion of the entire operation cost.
Furthermore, Wormhole's concurrency control allows updates on the MetaTrieHT to impose
minimal constraint on lookups' concurrency.

\begin{figure}[!t]
  \centering
  \includegraphics[width=0.9\columnwidth]{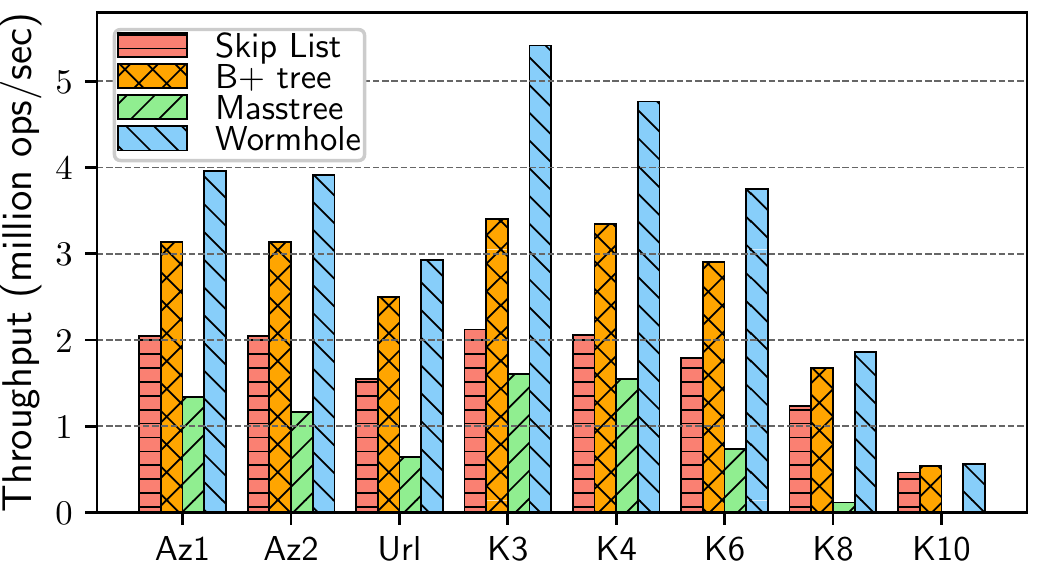}
  \caption{Throughput of range lookups}
  \label{fig:wh-scan}
\end{figure}

To compare Wormhole with other indexes on range operations,
we randomly select a search key and search for following (up to) 100 keys starting from the search key.
As range-scan is not implemented in the ART source code, it is omitted in this experiment.
The results for various keysets are shown in Figure~\ref{fig:wh-scan}.
In the range search much of the operation time is spent on sequentially scanning of a sorted list.
The performance advantage of Wormhole on reaching the first search key is dwarfed.
As a result, Wormhole's throughput improvement is reduced
(improvement of 1.05$\times$ to 1.59$\times$ over B+ tree).
However, as Masstree stores all keys in a trie structure, range query is much more expensive due to
its frequent pointer chasing on the trie, which leads to its much lower throughput than the other indexes.

\section{Related Work}
\label{sec:wh-related}
Comparison-based ordered indexes are commonly used as in-memory index of popular SQL and NoSQL databases,
such as B-tree (or B+ tree) in LMDB~\cite{lmdb} and MongoDB~\cite{MongoDB},
and skip list in MemSQL~\cite{MemSQL} and LevelDB~\cite{leveldb}.
Because their lookup cost is bounded by $O(\log N)$ the efforts on improving their lookup performance
are mainly focused on improvement of parallelism and caching efficiency.
For example, Bw-tree enables latch-free operations of B+ tree to improve lookup
efficiency on multi-cores~\cite{LLS13}.
FAST leverages architecture-specific knowledge to optimize B+-tree's
layout in the memory to minimize cache and TLB misses~\cite{KCS10}.
Many studies have proposed to use hardware accelerators, such as GPU,
to improve index lookups without changing the underlying data structure~\cite{HY11,ZWY15,KGP13,HSP13,SJ16}.
Wormhole takes a new approach to fundamentally reduce its asymptotic cost
to $O(\log L)$.
In addition to its algorithmic improvement,
Wormhole is further strengthened by a series of implementation optimizations.

Trie has been proposed to achieve a lookup cost lower than those of the comparison-based indexes.
ART adaptively changes the size of each trie node to minimize the space usage of the trie structure~\cite{LKN13}.
However with a small fanout (256 in ART) the $O(L)$ lookup cost can be significant for long keys.
Masstree enables a very high fanout ($2^{64}$) by using a B+ tree at each trie node~\cite{MKM12}.
Accordingly, Masstree's lookup cost on the trie structure is practically reduced to 1/8 of that in ART.
However, with the high fanout a trie node may have to be represented by a large B+ tree,
which makes access on this trie node slow and offsets the benefit of having reduced trie height.
Wormhole's lookup efficiency is less sensitive to  key length as it has a $O(\log L)$ lookup cost.
Using large leaf nodes to host keys and a small trie to manage the anchors,
Wormhole's space efficiency is much better than a trie.

Caching can effectively improve index lookup for workloads of strong locality.
For example, SLB uses a small cache to reduce the lookup cost for frequently
accessed data~\cite{WNJ17}.
However, caching is not effective for accessing of cold data.
Wormhole improves the index structure which can reduce DRAM accesses for workloads of little locality.
B$\varepsilon$-Tree is a B-tree-like index which allocates a buffer at each internal node
to reduce the high write amplification of B Tree~\cite{BF03}.
However the use of buffers incurs an additional overhead for lookups.
Similarly, FloDB uses a hash table as a buffer ahead of a skip list in LevelDB to service write requests,
which can remove the expensive skip-list insertion out of the critical path~\cite{BGT17}.
FloDB's hash table needs to be fully flushed upon serving a range operation,
which can impose long delays for range queries.
Wormhole has a low lookup cost which benefits both read and write operations.
By quickly identifying a leaf node for write operations, and using hashed keys in the sorting,
write operations in Wormhole has a consistently low cost.

In addition to using fine-grained locks,
many synchronization approaches have been proposed for efficient access of shared data structures.
MemC3~\cite{FAK13} and Masstree~\cite{MKM12} use version numbers to enable lock-free
access for readers.
Atomic operations, such as CAS and LL/SC, have been extensively used to implement
lock-free lists and trees~\cite{FR04,NM14,BP12,BFG05}.
RCU has been extensively used for read-dominant data structures~\cite{urcu,mckenney2013rcu}.
Other approaches, such as transactional memory and delegation techniques,
have been extensively studied~\cite{ST95,HM93,REB17,HIS10}.
We employ fine-grained locking, RCU, and version numbers to enable an efficient thread-safe Wormhole index
that is only slightly slower than the thread-unsafe Wormhole.
While there could be many other choices for more efficient concurrency control on Wormhole,
we leave it for future work.

\section{Conclusion}
\label{sec:wh-conc}

To the best of our knowledge, Wormhole is the first ordered key-value index achieving
the $O(\log L)$ lookup cost,
which is better than the $O(\log N)$ or $O(L)$ cost of other ordered indexes,
assuming key length $L$ much smaller than key count $N$.
The reduced asymptotic cost makes Wormhole capable of delivering quick access to KV items,
especially in challenging scenarios where the index manages a very large number of items with long keys.
Extensive evaluation demonstrates that Wormhole can improve index lookup throughput
by up to 8.4$\times$, 4.9$\times$, 4.3$\times$, and 6.6$\times$,
compared with skip list, B+ tree, ART, and Masstree, respectively.
Meanwhile, Wormhole's performance with other operations, including insertion, deletion,
and range query, is also higher than or comparable to other indexes.
Its space demand is as low as that of B+ tree.
The source code of an implementation of the Wormhole index is publicly available at
\mbox{texttt{https://github.com/wuxb45/wormhole}}.

{\footnotesize
\bibliographystyle{abbrv}
\bibliography{01-master}
}
\end{document}